\documentstyle[aps,prl,preprint]{revtex}
\begin{document}
%\vspace*{-62pt}

\date {SUSX-TH-95/3-3 $~~~~$  Fermilab-Pub-95/021-A$~~~$ DART-HEP-95/01 $~~~~$
hep-ph/9503217}

\vspace{0.5in}
\title{Oscillons: Resonant Configurations During Bubble Collapse}

\vspace{1.cm}

\author{E. J. Copeland$^{1)}$, M. Gleiser\thanks{NSF Presidential Faculty
Fellow. On leave from Department of
Physics and Astronomy, Dartmouth College, Hanover NH 03755.
}$^{2)}$, and H.-R. M\"uller$^{3)}$}

\vspace{1.0cm}

\address{ $^{1)}$ School of Mathematical and Physical Sciences, University
of Sussex \\
Brighton BN1 9QH, UK}

\address{ $^{2)}$ Nasa/Fermilab Astrophysics Center\\
Fermi National Accelerator Laboratory\\
P.O.Box 500, Batavia, IL 60510, USA}

\address{ $^{3)}$ Department of Physics and Astronomy, Dartmouth College\\
Hanover, NH 03755, USA}

\maketitle

\vspace{1.cm}

\begin{abstract}
\baselineskip 16pt
Oscillons are localized, non-singular, time-dependent, spherically-symmetric
solutions of nonlinear
scalar field theories which, although unstable, are {\it extremely} long-lived.
We show that they naturally appear during the collapse of subcritical bubbles
in models with symmetric and asymmetric double-well potentials.
By a combination of analytical and numerical work we explain several of their
properties, including the conditions for their existence, their longevity, and
their final demise. We discuss several contexts in which we
expect oscillons to be relevant. In particular, their nucleation during
cosmological phase transitions may have wide-ranging consequences.

%\vspace{1.cm}

\end{abstract}
\newpage

\baselineskip 24pt
\section{Introduction}

The search for static, localized, non-singular solutions of nonlinear
field theories has by now a long history \cite{SOLITONS}. In (1+1)-dimensions,
it is possible to find exact static solutions to the nonlinear Klein-Gordon
field equations for certain interacting potentials,
such as the kink solutions of sine-Gordon or $\phi^4$ models. For a
larger number of spatial
dimensions, Derrick's theorem forbids the existence of static solutions for
models involving only real scalar fields \cite{DERRICK}.
There are several ways to circumvent
Derrick's theorem, by invoking more complicated models with two or more
interacting fields. Well-known examples include topological defects such
as the 't Hooft-Polyakov monopole or the Nielsen-Olesen vortices
\cite{RAJARAMAN}.
Topological conservation laws guarantee the stability
of these configurations.

It is also
possible to find localized time-dependent but non-dissipative
solutions of nontopological nature, the so-called
nontopological solitons \cite{NTSs}. The simplest model of a nontopological
soliton in the context of renormalizable theories
has a complex scalar field quadratically
coupled to a real scalar field with quartic potential.
The stability of the
configuration comes from the conserved
global charge $Q$ carried by the complex field which
is confined within a spherically-symmetric domain formed by the real scalar
field. One can show that for $Q$ larger than a critical value,
the energy of the configuration is smaller than the energy of $Q$ free
particles. There has been a recent upsurge of interest on nontopological
solitons due to their potential relevance to cosmology and astrophysics
\cite{NTSCOS}. If one waives the requirement of renormalizability, it is
possible to find nontopological solitons for models with
a single complex scalar
field, by invoking, e.g., a $\phi^6$ term in the potential. These are the
so-called $Q$-ball
solutions discovered by Coleman and collaborators \cite{QBALL}.

In the present work we will go back to the simple models involving only
a self-interacting real scalar field and study the properties of
{\it time-dependent} spherically-symmetric solutions. Due to the constraint
imposed by Derrick's theorem, these configurations have been somewhat
overlooked in the literature (but not completely, as we will discuss below).
Why should anyone bother with solutions which are known to be unstable?
One possible answer is that instability is a relative concept, which only
makes sense in context, that is, when the lifetime of a given configuration
is compared with typical time-scales of the system under study. Thus,
unstable but long-lived configurations may be relevant for systems with short
dynamical time-scales. Another answer is that a detailed study of these
configurations can greatly clarify dynamical aspects of nonlinearities in field
theories and the r\^ole they play in several phenomena, ranging from
nonlinear optics to phase transitions both in the laboratory and in
cosmology \cite{TEXTURES}.

One of the motivations for studying the evolution of
unstable spherically-symmetric configurations comes from the work of Gleiser,
Kolb, and Watkins on the r\^ole subcritical bubbles may play in the
dynamics of weak first order phase transitions \cite{GKW}. Considering models
with double-well potentials in which the system starts localized
in one minimum, these authors proposed that for sufficiently weak transitions
correlation-volume bubbles of the other phase could be thermally
nucleated, promoting an effective phase mixing between the two available
phases even before the critical temperature is reached from above.
This could have
important consequences for models of electroweak baryogenesis which
rely on the usual homogeneous nucleation mechanism \cite{GK}. However, Gleiser,
Kolb, and Watkins did not include the shrinking of the bubbles in their
estimate of the fraction of the volume occupied by each of the two phases,
leading some authors to question their results \cite{CRITICS}. Since then,
Gleiser and Gelmini included the shrinking of the bubbles into the original
estimates, concluding that for sufficiently weak transitions
subcritical bubbles are indeed
nucleated at a fast enough rate to cause substantial phase mixing \cite{GG}.
Although an improvement, the modeling used to describe
the bubble shrinking was still too simplistic,
as it assumed that the bubbles just shrunk with constant velocity.

The evolution of spherically-symmetric unstable solutions of the nonlinear
Klein-Gordon equation was originally studied numerically
in the mid-seventies by
Bogolubsky and Makhankov \cite{PULSON1}. Using a quasiplanar initial
configuration for the bubbles (that is, a tanh$(r-R_0)$ profile, with
$R_0$ the initial radius), these authors discovered that for a
certain range of initial radii the bubble evolution could be described in three
stages; after radiating most of its initial energy the bubble settled into
a regime which was quite long-lived, with a lifetime which
depended on the initial radius. The bubble then disappeared by quickly
radiating away its remaining energy.
These configurations were called ``Pulsons''
by these authors, due to the pulsating mechanism by which they claimed the
initial energy was being radiated away.
Their results were recently rediscovered
and refined by one of us \cite{PULSONMG}. After a more detailed analysis of
these configurations, it became clear that their most striking
feature was not the pulsating mechanism by which bubbles radiate
their initial energy, but the rapid oscillations of the
field's amplitude at the core of the configuration during the pseudo-stable
regime, in a manner somewhat analogous to resonant breathers in
kink-antikink scattering \cite{BREATHERS}.
In fact, it was realized
that during the pseudo-stable regime almost no energy is radiated away
and the radial pulsation is actually quite small in amplitude. Hence the name
``Oscillon'' was proposed instead. It was also shown that these configurations
appear both in symmetric and asymmetric potentials, are stable against small
radial perturbations, and have lifetimes far
exceeding naive expectations. However, not much else has been done in order
to explore the properties of these configurations. Other works on this topic
were concerned in establishing the existence of these solutions for other
potentials, such as the sine-Gordon and logarithmic potentials, different
symmetries, and somewhat limited stability studies \cite{PULSON2}.

By a combination of analytical and numerical methods, we will shed some light
on the properties of these configurations (henceforth oscillons).
We will establish the conditions for
their existence, the reason for their longevity, and clarify their final
collapse. Armed with a better understanding of their properties, we will also
be able to suggest several situations where we believe oscillons
can be of importance.

The rest of this paper is organized as follows. In the
next Section we will set up the general formalism and obtain the exact solution
of the spherically-symmetric linear Klein-Gordon equation. As expected,
in the linear case no oscillons appear, with bubbles quickly decaying away. We
obtain the time-scale in which this decay occurs in order to later
compare it to the case when nonlinearities are present. In Section 3 we
present the numerical results that establish several of the key properties
of oscillons for symmetric double-well potentials.
Guided by these results, in Section 4 we present analytical
arguments to explain why there is a minimum initial radius for bubbles
to settle into the oscillon stage,
why some oscillons live longer than others, and how oscillons finally
disappear. In Section 5 we extend the numerical analysis of
Section 3 to asymmetric double-well
potentials, showing how the lifetime of oscillons is
sensitive to the amount of asymmetry between the two minima.
Here one must be careful to set the initial radius to be smaller than the
critical radius, as bubbles with radii larger than critical will
grow. As in the symmetric potential case there are no critical bubbles, we
can say that we are studying the evolution of subcritical bubbles in
symmetric and asymmetric potentials. Oscillons are thus a possible stage in the
evolution of subcritical bubbles toward their demise. In Section 6 we
discuss several possible situations in which these configurations will play
an important r\^ole. Although we focus mainly
on cosmological phase transitions, some
of our arguments apply equally well to phase transitions in the laboratory.
We conclude in Section 7 with a summary of our results and an outlook to
future work.

\section{Preliminaries}

In this Section we introduce the notation and some definitions which will be
useful later on. We also present the exact solution for the evolution of
a ``Gaussian-shaped'' bubble ({\it i.e.} with $\phi(r,t=0)\sim {\rm exp}
[-r^2/R^2]$) in the linear regime.

\subsection{General Formalism}

The action for a real scalar field in (3+1)-dimensions is
\begin{equation}
S[\phi] = \int d^4x \left[ {1\over 2} (\partial_{\mu} \phi)
(\partial^{\mu} \phi) - V_{\rm S(A)}(\phi) \right ] ~~,
\end{equation}
\noindent where the subscripts S and A stand for symmetric (SDWP)
and asymmetric (ADWP) double-well
potentials, given respectively by,
\begin{equation}\label{potsdwp}
V_{\rm S}(\phi) = {\lambda\over 4} \left( \phi^2 -
 {m^2\over\lambda} \right)^2
\end{equation}
\noindent
and,
\begin{equation}\label{potadwp}
V_{\rm A}(\phi) = {m^2\over 2}\phi^2 -
{{\alpha_0 m}\over 3} \phi ^3 + {\lambda\over 4}  \phi^4 ~~.
\end{equation}
\noindent
Note that the coupling constants $\lambda$ and $\alpha_0$ are dimensionless. A
solution $\phi({\bf x},t)$ to the equation of motion,
\begin{equation}\label{fullEoM}
\partial^2\phi/\partial t^2 -
\nabla^2 \phi = - {\partial V(\phi)\over \partial \phi} ~~,
\end{equation}
\noindent
has energy
\begin{equation}
E[\phi] = \int d^3x \left[ {1\over2} (\partial\phi/\partial t)^2 +
{1\over 2} (\nabla \phi) ^2 + V(\phi) \right] ~~.
\end{equation}

We will restrict our investigation to spherically-symmetric configurations.
In this case
it proves convenient to introduce dimensionless variables,
$\rho=rm,~\tau=tm,~{\rm and}~\Phi={{\sqrt{\lambda}}\over m}\phi$. The nonlinear
Klein-Gordon equation is,
\begin{equation}\label{eqmotsp}
{{\partial^2\Phi}\over {\partial\tau^2}}-{{\partial^2\Phi}
\over {\partial \rho^2}}
-{2\over {\rho}}{{\partial \Phi}\over {\partial \rho}}  =  \left\{
\begin{array}{ll}\Phi - \Phi^3 & \mbox{(SDWP)} \\
-\Phi + \alpha\Phi^2 -\Phi^3 & \mbox{(ADWP)}
\end{array}
 \right.
\end{equation}
\noindent
where $\alpha=\lambda^{-1/2}\alpha_0$. Note that for the SDWP
the two minima are located at $\Phi_0=-1$ and $\Phi_+=1$.
For the ADWP (with $\alpha\geq 2$), the minima are at
$\Phi_0=0$ and $\Phi_+={{\alpha}\over 2}\left [1+
\left (1 - {4\over {\alpha^2}}\right )^{1/2}\right ]$.
Requiring $\Phi_+$ to be the global
minimum implies $\alpha^2> 9/2$. For $\alpha^2=9/2$ the
two minima are degenerate with $\Phi_+=\sqrt{2}$.
This value will be important later.

We are interested in following the evolution of unstable
spherically-symmetric configurations of initial radius $R_0$, (from now on we
call these initial
configurations subcritical bubbles both for the SDWP and the ADWP)
which can be thought of as being
localized fluctuations about the global vacuum $\Phi_0$.
Thus, we must measure the rate at which the
initially localized energy is radiated away
as the subcritical bubble relaxes to the global vacuum.
This can be done by surrounding the initial configuration
with a sphere of sufficiently large radius, $R_s\gg R_0$,
and measuring the flow of energy through the surface of the sphere.
The evolution of subcritical bubbles is obtained by
solving the nonlinear Klein-Gordon equation numerically. We define the
subcritical bubble's kinetic, surface, and volume energies, respectively, by
\begin{equation}\label{energies}
E_k  =  2 \pi \int_0^{R_s} \rho^2 \dot\Phi^2 \, d\rho,\;
E_s  =  2 \pi \int_0^{R_s} \rho^2 (\Phi')^2 \, d\rho,\;
E_v  =  4 \pi \int_0^{R_s} \rho^2  V(\Phi) \, d\rho ~,
\end{equation}
\noindent
where a prime denotes derivative with respect to $\rho$.
The bubble's total energy is thus,
\begin{equation}
E_b(\tau)=E_k(\tau)+E_s(\tau)+E_v(\tau)~~.
\end{equation}

In order to solve the nonlinear Klein-Gordon equation we will impose the
following boundary conditions,
\begin{equation}\label{bc}
\Phi(\rho\rightarrow \infty,\tau) = \Phi_0,\; \Phi'(0,\tau) = 0,\;
\dot\Phi(\rho,0) = 0\, .
\end{equation}
\noindent
The first condition guarantees that the bubble approaches the vacuum at
$\Phi_0$ at spatial infinity.
The second condition imposes regularity at the
origin, while the last condition states that the bubbles start their evolution
at rest. These conditions must be supplemented by the initial
profile of the bubble. We will investigate both `Gaussian' and `tanh' bubbles
which we write as
\begin{eqnarray}\label{initg}
\mbox{Gaussian:  } &\Phi(\rho,0)& =
(\Phi_c - \Phi_0)\, e^{-\rho^2/R_0^2} + \Phi_0\\
\label{initt}
\mbox{tanh:  } &\Phi(\rho,0)& ={1\over 2}\left[ (\Phi_0 - \Phi_c)\,
\tanh (\rho - R_0) +
\Phi_0 + \Phi_c \right]~,~~R_0\gg 1~.
\end{eqnarray}
\noindent
$\Phi_c$ is the value of the field at the bubble's core, which we may or not
take as being the other minimum of the potential, $\Phi_+$.
If we do, the bubble can
be interpreted as being a field configuration of initial
linear size $\sim 2R_0$ which interpolates between the two vacua. As we will
see later, it is not necessary to set $\Phi_c=\Phi_+$ in order to have
subcritical bubbles relaxing into oscillons during their evolution. We can
now move on to study the evolution of subcritical bubbles in the linear regime.

\subsection{Bubble Evolution in the Linear Regime}

As a first application of the above formalism, we will investigate the
evolution of Gaussian bubbles in the linear regime.
We choose as the linear potential,

\begin{equation}
V_{\rm L}(\Phi)=\left (\Phi+1\right )^2~~,
\end{equation}
\noindent
as it has a minimum at $\Phi_0=-1$ with the same curvature as the SDWP. The
Klein-Gordon equation has a trivial solution $\Phi(\rho,\tau)=-1$. Separation
of
variables with a constant $-k^2$ allows us to write
$\Phi(\rho,\tau)=-1 +R(\rho){\rm exp}[\pm i\sqrt{k^2+2}\tau]$,
with the radial function
$R(\rho)$ obeying,

\begin{equation}
R'' + {2 \over {\rho}} R' + k^2 R = 0~~.
\end{equation}

\noindent
This equation has solutions which are linear combinations of
${\sin k \rho \over \rho},~ {\cos k \rho \over \rho}$.
Since ${\cos k \rho \over \rho}$ is
singular at the origin we write the general solution as
\begin{equation}\label{presol1}
\Phi (\rho,\tau) = -1 + \int_{0}^{\infty} dk\: b(k) {\sin k \rho \over \rho}
\left[ \cos (\sqrt{k^2 + 2}\, \tau) + a(k) \, \sin (\sqrt{k^2 + 2}\, \tau)
 \right]\, .
\end{equation}
\noindent
The boundary conditions are, writing $\Phi_c\equiv 2q_0-1$, $q_0$ an arbitrary
constant,
\begin{eqnarray}
\label{init2}\Phi (\rho,\tau=0) & = & 2 q_0 \, e^{-\rho^2/R_0^2} - 1 \\
\label{faraway}\Phi (\rho \to \infty , \tau) & = & -1 \\
\label{zeroder2}\Phi' (\rho=0,\tau) & = & 0 \\
\label{time} \dot\Phi(\rho,\tau=0) &= &0 ~.
\end{eqnarray}
\noindent
Eq.~\ref{faraway} is trivially satisfied. Eq.~\ref{init2} determines
$b(k)$,
\begin{equation}
2 q_0 \, e^{-\rho^2/R_0^2} =
\int_{0}^{\infty} dk\: b(k) {\sin k \rho \over \rho}~.
\end{equation}
\noindent
Taking the sine transform we can write,
\begin{equation}
b(k) = {4 q_0 \over \pi} {\rm Im} \left[
\int_{0}^{\infty} d\rho\: \rho\, e^{-\rho^2/R_0^2} e^{i k \rho} \right]~.
\end{equation}
\noindent
The integral can be easily done and we obtain,
\begin{equation}
b(k)={q_0 R_0^3\over \sqrt{\pi}}\, k\, e^{-R_0^2 k^2/4}~.
\end{equation}

Regularity at the origin is also guaranteed, as $\left [{{sin k\rho}\over
{\rho}}
\right ]^{\prime}$ vanishes as $\rho \to 0$. Choosing the bubble to
start at rest implies that $a(k)$ vanishes. Thus, the final solution satisfying
all boundary conditions is,
\begin{equation}\label{linsol}
\Phi (\rho,\tau)  = -1 + {q_0 R_0^3\over \sqrt{\pi}} \int_{0}^{\infty} dk\:
 k\, e^{-R^2_0 k^2/4}\, {\sin k \rho \over \rho}
\, \cos (\sqrt{k^2 + 2}\, \tau)
 \, .
\end{equation}

In Figure 1 we show a plot of this solution, for initial amplitude $q_0=1$
and radius $R_0=3$. The bubble performs damped oscillations as it decays
into $\Phi_0=-1$.

It is instructive to investigate the behavior of the bubble's core with time,

\begin{equation}
\Phi (\rho=0,\tau) = -1 + {q_0 R_0^3\over \sqrt{\pi}} \int_{0}^{\infty} dk\:
 k^2\, e^{-R_0^2 k^2/4}\, \cos (\sqrt{k^2 + 2}\, \tau)\, .
\end{equation}

\noindent
The integral is dominated by small values of $k$,
$k\lesssim 2R_0^{-1}$.
Thus, we can
approximate the argument of $\cos (\sqrt{k^2+2}\tau)$ for
$R_0\gtrsim 2$ and write,
\begin{equation}
\Phi (\rho=0,\tau) = -1 + {q_0 R_0^3\over \sqrt{\pi}} \mbox{Re} \left[
e^{i \sqrt{2} \tau}  \int_{0}^{\infty} dk\:  k^2\, e^{-R_0^2 k^2/4}\,
e^{i \sqrt{2} \tau k^2/4} \right]~~.
\end{equation}
\noindent
Performing the integral we obtain
\begin{equation}
\Phi (\rho=0,\tau) = -1 + {2 q_0\over {(1 + 2 \tau^2 / R_0^4)^{3/4}}} \,
\cos \left( \sqrt{2} \tau + {3\over 2} \tan ^{-1} \left(
\sqrt{2} \tau \over R_0^2 \right) \right )~~.
\end{equation}
Thus, the amplitude at the core decays as $\tau^{-3/2}$,
while the frequency becomes constant
for $\tau\gtrsim R^2_0/\sqrt{2}$. The envelope of the
core's
amplitude decays to $1/e$ of its initial value above $\Phi_0=-1$
in a time (units restored)
%%%NOTE that R_0 is defined as a dimensionless number!!%%%%

\begin{equation}
t_{1/e}\simeq
1.18 R^2_0 m^{-1}~.
\end{equation}
\noindent In Fig.~2 we compare the above analytical approximation with the
numerical solution of the Klein-Gordon equation (more details later). The
excellent agreement gives support to the accuracy of the numerical
methods used.

\section{Evolution of Subcritical Bubbles in SDWP: Numerical Results}

In this and the next Section we will restrict our analysis to bubbles in
SDWP. Section 5 will deal with bubble evolution for ADWPs. The equation
of motion is
\begin{equation}
\ddot\Phi - \Phi'' -{2\over {\rho}} \Phi' = \Phi - \Phi^3~~,
\end{equation}
\noindent
with boundary conditions given by Eqs.~\ref{init2} -- \ref{time}. (For
tanh bubbles or any other initial bubble profile,
just replace Eq.~\ref{init2} by the appropriate choice.) This equation was
solved numerically using a finite difference scheme fourth order accurate in
space and second order accurate in time. The radial dimension of the
two-dimensional grid moved outwards
with the speed of light in order to avoid any
radiation from being reflected on the lattice boundary and thus
interfering with the
bubble's evolution within the grid (dynamically increasing simulation lattice).
The alternative, a sufficiently long but static
grid, is extremely time-consuming for
long-lived oscillons.
The resolution was typically set to $\Delta \rho = 0.1$ and $\Delta \tau =
0.05$;
the total energy
$E_b + 4 \pi \int_{R_s}^{\infty} \rho^2  \left[ {1\over2} \dot\Phi^2 +
{1\over 2} (\Phi') ^2 + V(\Phi) \right] d\rho$ is then conserved
throughout the evolution to better than one part in $10^3$. Additionally,
high resolution experiments ($\Delta \rho = 0.01,\;\Delta \tau = 0.005$)
produced
the same results, with an energy
conservation of one part in $10^5$.
We used $R_s = 10$ for the SDWP, and $R_s = 15$ for the ADWP.
For reasons that will be made clear soon, we were not interested in bubbles of
large initial radius.

Figs.~3a and 3b show the energy of Gaussian and tanh bubbles
for several initial radii $R_0$. (More examples can be found in
Ref.~\cite{PULSONMG}.) In all the examples we took $\Phi_c=+1$; the bubble
interpolates between the two vacua.
It is clear that the evolution of the bubbles is very
sensitive to the value of $R_0$. An extensive investigation showed that
Gaussian
bubbles with $R_0\lesssim 2.4$ and
$R_0\gtrsim 4.5$
quickly disappear, radiating
their initial energies to infinity. However, bubbles with
$2.4\lesssim R_0 \lesssim 4.5$,
settle into a period of long-lived stability where practically no energy
is radiated away. This stage in their evolution, which we call the
oscillon stage, can have a duration approaching $10^3 - 10^4m^{-1}$, which
is remarkably large compared to both short-lived bubbles and to
the typical time-scales found for the linear potential. Although the
range of values for $R_0$ which fall into an oscillon stage
is sensitive to the initial profile of the configuration,
the same results are obtained for other initial bubbles, such as tanh bubbles.
This supports our previous claim that oscillons can be
viewed as a possible stage during the evolution of subcritical bubbles;
after shedding a sufficient fraction of their initial energy,
the subcritical bubbles enter the oscillon stage which is characterized by
an energy with a nearly constant value of $\sim 43 m/\lambda$,
regardless of their initial radius.

The core value of the field, $\Phi(0,\tau)$,
performs anharmonic oscillations as shown in Fig.~4a. In Fig.~4b
we show a sequence
of snapshots of an oscillon.
Except when $\Phi(0,\tau)\simeq -1$,
an oscillon configuration is very well approximated by a half-Gaussian. We
define the effective radius of a localized field configuration by,

\begin{equation}\label{radius2}
R_{\rm eff}(\tau) = {{\int_{0}^{R_s} \rho^3  \left[
{1\over 2} (\Phi') ^2 + V(\Phi) \right] d\rho}\over{\int_{0}^{R_s}
\rho^2  \left[
{1\over 2} (\Phi') ^2 + V(\Phi) \right] d\rho}}~~,~~ R_{\rm eff} \ll R_s~.
\end{equation}

\noindent
In Fig.~5 we show the evolution of this radius for several bubbles
during the oscillon stage. The divergence at the end is spurious, signaling
that no energy is left within the sphere
(the denominator of Eq.~\ref{radius2}).
It is clear that the effective radius
of an oscillon is approximately constant, with variations which are smaller
than $20\%$ about a mean of $R_{\rm osc}\simeq 2.8 - 3.0$.
This justifies the name given to these configurations:
An oscillon is a localized, time-dependent
field configuration with nearly
constant radius and energy which is characterized by
anharmonic oscillations of the
field amplitude about the global vacuum.

In order to stress the remarkable longevity of oscillons we show in Figs.~6a
and 6b the lifetime as a function of initial radius and energy, respectively.
For Gaussian bubbles the
longest living oscillon, with $\tau_l\simeq 7.4 \times 10^3$,
comes from an initial bubble of radius $R_0=2.86$. For tanh bubbles, the
longest living oscillon comes from an initial bubble of radius $R_0=3.08$,
with lifetime $\tau_l\simeq 4.4\times 10^3$.
In Fig.~7 we show the detailed dependence of lifetime as a function of
radius for Gaussian bubbles about the peak at $R_0=2.86$.
(Lifetimes are accurate to within $5\%$.)

So far we have restricted our investigation to bubbles that
interpolate between the two vacua. This is {\ it not}
a necessary condition for the
existence of oscillons although, of course, it is sufficient.
As long as the initial value of the field at the
bubble's core
probes the nonlinearities of the potential and the initial radius is
within the correct range (which varies with initial amplitude),
oscillons can exist. We will give an analytical
argument for this in the next Section. For now, we will just provide numerical
evidence for this fact. In Fig.~8 we show
a plot of lifetime for different core values. Clearly, no oscillon can
develop if the initial energy is below the plateau energy. Also, we find
that no oscillon develops if $\Phi_c \leq
\Phi_{\rm inf}$, where $\Phi_{\rm inf} = -1/\sqrt{3}$
is the inflection point closest to $\Phi_0$.
Thus, we arrive at
the {\it sufficient conditions} for the existence of oscillons:
i) the value of the field at the bubble's core must be above the inflection
point, and
ii) the initial bubble's energy must be above the plateau energy. Conditions
i) and ii) fix the value of $R_0$ for a given initial bubble to evolve into
an oscillon.

This concludes the presentation of our numerical results. In the next Section
we will provide semi-analytical arguments to elucidate some of the properties
of these configurations.

\section{Properties of Oscillons}

{}From the results of the previous Sections, it is clear that
there are four main questions
concerning the oscillons. First, why only bubbles with an initial
radius above a
certain value develop into oscillons, and how this value depends on the
initial amplitude of the field at the bubble's core. Second, why certain
oscillons live longer than others. Third, what is the mechanism responsible
for the oscillon's final collapse.
And finally, why above a maximum
initial bubble radius no oscillons are possible. In this Section we address
the first three questions. Work on the fourth question is in progress.

In order to treat these questions analytically, we make
use of the fact that independently of the initial bubble profile, an oscillon
is very well approximated by a half-Gaussian. Even though the Klein-Gordon
equation implies that $\Phi(\rho\to \infty ,\tau)\sim {\rm exp}[-\rho ]$, the
difference turns out to be sufficiently small in practice to justify our
approximation. In a sense, the tail matters little to the dynamical properties
of the configurations.
The agreement of our analytical arguments with the numerical
results should convince the reader of this fact. We model the
oscillon by writing, for the SDWP,

\begin{equation}\label{ansatz}
\Phi(\rho,\tau) = 2q(\tau){\rm exp}\left [-\rho^2/R^2(\tau)\right ] -1 ~~.
\end{equation}

\noindent With this {\it ansatz} we have effectively reduced the field theory
problem to two degrees of freedom, the amplitude at the core $
\Phi_c(\tau)\equiv 2q(\tau)-1$, and the
radius $R(\tau)$. This problem is still quite complicated to treat analytically
due to the nonlinear coupling between the two degrees of freedom. Further
simplification is guided by the numerical investigation, which
showed that the effective radius of the oscillon remains practically
constant, with oscillations about its  mean value
of order $20\%$ or less. Thus, as a
first step, we will keep the radius constant, and treat only the amplitude
at the core as an effective degree of freedom. Strong as it may seem, this
simplification will allow us to extract several important results concerning
the observed numerical behavior of these configurations, as we show
in the next subsections. We are currently investigating the consequenses of
keeping both degrees of freedom $q(\tau)$ and $R(\tau)$.

The above model for the oscillon
still misses one important ingredient;
it does not include radiation of the bubble's
energy to infinity.
The justification for neglecting this
lies in the fact that oscillons hardly radiate.
By excluding radiation it is possible to analytically
integrate the energy over the whole space. Using the
definitions in Eq.~\ref{energies} we obtain,
for the kinetic, surface and volume energies, respectively,
\begin{equation}\label{enterms}
E_k={{\pi\sqrt{2\pi}}\over 2}R^3\dot q^2,~E_s={{3\pi\sqrt{2\pi}}\over 2}Rq^2,~
E_v=\pi\sqrt{2\pi}R^3\left (q^2-{{4\sqrt{6}}\over 9}q^3+{{\sqrt{2}}\over 4}q^4
\right )~.
\end{equation}

\subsection{Existence of Oscillons: Lower Bound on the Initial Radius}

{}From Fig.~6a it is clear that there is a lower bound on the initial value of
the bubble radius so that it relaxes into an oscillon during
its collapse. Since from our previous discussion we know that oscillons are
a product of the nonlinearities in the system, this result suggests that for
small enough initial radii the nonlinearities are ineffective to trigger the
resonant behavior responsible for the oscillon's longevity. That this is
the case can be shown by studying the effective potential controlling the
behavior of the amplitude $q(\tau)$. Using the above {\it ansatz} with
constant radius, the energy of the configuration $E = E_k + E_s + E_v$
can be written as the energy
of a particle of unit mass with a potential $V(q)$,

\begin{equation}\label{oscener}
{E\over {16\pi A}}={1\over 2}\dot q^2+V(q),~~V(q)=
\left (1+{B\over A}\right )q^2
-{C\over A}q^3+{D\over A}q^4~~,
\end{equation}

\noindent
where $A={{\sqrt{2\pi}}\over {16}}R^3,~B={{3\sqrt{2\pi}}\over {32}}R,~C={{
\sqrt{3\pi}}\over {18}}R^3,~{\rm and}~D={{\sqrt{\pi}}\over {32}}R^3$
follow from Eq.~\ref{enterms}. The
potential $V$ has only one minimum at $q=0$, ($\Phi=\Phi_0=-1$,
the global vacuum), about which the amplitude performs anharmonic
oscillations.

It is the energy localized within a small region surrounding the bubble
that may (or not) feed the nonlinear growth of the modes ultimately
responsible for the appearance of the oscillon during the collapse of the
bubble. This lends further support to the  above {\it ansatz}
neglecting radiation. Thus,
the equation of motion
for the amplitude $q(\tau)$ is,

\begin{equation}
\ddot q = -2\left (1+{B\over A}\right )q + 3{C\over A}q^2 - 4{D\over A}q^3~~.
\end{equation}

\noindent
Writing $q(\tau)={\bar q}(\tau)
+\delta q(\tau)$, the linearized equation satisfied by the
fluctuations $\delta q(\tau)$ is,

\begin{equation}\label{omegaeq}
\ddot{\delta q} = -\omega^2({\bar q},R) \delta q,~~\omega^2({\bar q},R)\equiv
3\sqrt{2}{\bar q}^2-8{{\sqrt{6}}\over 3}{\bar q}
+\left (2 +{3\over {R^2}}\right )~~,
\end{equation}

\noindent
where we have substituted the numerical values of the constants $A,~B,~C,$ and
$D$ in the expression for the frequencies $\omega^2({\bar q},R)$. Note that
$\omega^2({\bar q},R)$ is simply the curvature of the potential dictating the
dynamics of the amplitude $q(\tau)$.
As the bubble radiates its energy away, the configuration decays
into the vacuum. However, for $\omega^2({\bar q},R)<0$, fluctuations
about ${\bar q}$
are unstable, driving the amplitude {\rm away} from its vacuum value. These
are the fluctuations which are mainly
responsible for the appearance of the oscillon.
In Fig.~9 we show a plot of the surface $\omega^2({\bar q},R)$.
It has one minimum
at ${\bar q}_{\rm min}\simeq 0.77$
(with location independent of $R$!),  where its value is
$\omega^2(q_{\rm min},R)\simeq -0.514 + 3R^{-2}$. Thus, only for $R> R_{\rm
min}
\simeq 2.42$, $\omega^2<0$ and fluctuations can grow. In other words,
{\it only for $R_0>R_{\rm min} \simeq
2.42$ are oscillons possible}. This lower bound on the value
of the radius agrees very well with our numerical results (see Fig.~6a). It is
independent of the initial amplitude of the bubble. All bubbles with
initial radius smaller than $R_{\rm min}$ will quickly collapse. (For core
amplitudes above $\Phi_c=1$, it is possible to decrease the initial radius
by about $20\%$ or so and still obtain oscillons.)

\subsection{Collapse of Oscillons}

The above analysis can also provide information about the final decay of
oscillons. For $R>R_{\rm min}$,
$\omega^2({\bar q},R)$ will be negative for amplitudes,

\begin{equation}\label{minamp}
{\bar q}_-~\leq ~{\bar q}~\leq ~{\bar q}_+~,~~{\bar q}_{\pm}=
{{4\sqrt{3}}\over 9}\left [1\pm
\left (1-{{9\sqrt{2}}\over 16}\left (1+{3\over {2R^2}}\right )\right )^{1/2}
\right ]~~.
\end{equation}

\noindent
Thus, for $R>R_{\rm min}$ there is a minimum value for the amplitude at
the core, shown in Fig.~10,
$\Phi_c^-(R)=2{\bar q}_-(R) - 1$,
below which the oscillon slips into the linear regime and quickly
decays. This result can be understood as follows:
As the bubble settles into the oscillon configuration with energy
given by the plateau energy $E\sim 43m/\lambda$ and radius
$R_{\rm eff}
\sim 2.8m^{-1}$, there is a maximum value for the amplitude of the field
at the core. This value is obtained from the formula for the static energy with
its value fixed at the plateau value and with radius $R\simeq R_{\rm eff}$,
and it
is $\Phi_c\sim 0.2$. The values of the field at the oscillon's core
obtained numerically are always marginally
within the allowed region which gives $\omega^2<0$; the oscillon survives while
fluctuations are unstable.
However, during the oscillon stage, energy is slowly being radiated
away, and thus the amplitude at the core is slowly decreasing while the
average value of the radius is slowly increasing (Fig.~5). From
Fig.~10 and the argument above, below a certain value for the amplitude at
the core the perturbations enter the linear regime and the oscillon decays.
A comparison between $\Phi_c^-(R_{\rm eff})$ and the numerical
values of the core's amplitude
at the last oscillation is given in Table 1. In interpreting these results, we
must keep in mind the crudeness of the analytical approximation used to
obtain $\Phi_c^-(R)$. Even so, at least for the longest living oscillons,
it is clear that during the last oscillation the
amplitude falls below $\Phi_c^-(R)$. A more detailed analysis shows that the
amplitude falls below $\Phi_c^-(R)$ during the last few oscillations, as the
configuration starts to approach the linear regime responsible for its
final demise.
For completeness, in Fig.~11
we show a phase-space portrait of the evolution of $\Phi_c(\tau)$
during the oscillon stage and its final collapse, for a bubble with initial
radius $R_0=3.0$. Clearly, the final
spiraling into $\Phi_0=-1$, typical of the linear
regime, occurs as the maximum core amplitude (for $\dot\Phi=0$) falls roughly
below $\Phi_c^-$.

\subsection{Lifetime of Oscillons}

A question which is of great interest is the determination of the oscillon's
lifetime as a function of the bubble's initial radius and core value. Although
we were unable to obtain an analytical expression for the lifetime, we do
understand why some oscillons live longer than others. Our argument is based
on the virial theorem for spherically-symmetric scalar field configurations,
which we derive next.

Multiplication of the equation of motion,
$\partial^2\phi/\partial t^2 - \partial^2\phi/\partial r^2 -
(2/r)\, \partial\phi/\partial r
= - {\partial V(\phi)\over \partial \phi}$
(cf.~Eq.~\ref{fullEoM}), by $4\pi r^2 \phi$
and integrating over $r$ gives, after integration by parts,
\begin{equation}
4 \pi \int_0^{\infty} r^2 \phi \ddot\phi\, dr
+ 4 \pi \int_0^{\infty} r^2 \phi'^2\, dr
+ 4 \pi \int_0^{\infty} r^2 \phi {\partial V \over \partial \phi} \, dr
= 0~~,
\end{equation}
where we assumed that ${\rm lim}_{r\to\infty}r^2\phi '=0$.
The second term is easily recognized as twice the total surface
energy. Performing a time averaging over one period, denoted by
\begin{equation}
\langle ~ \rangle \equiv {1\over T} \int_T dt
\end{equation}
we get
\begin{equation}
{4 \pi \over T} \int_0^{\infty} r^2 \, dr \int_T dt \phi \ddot\phi\
+ 2 \langle E_s\rangle
+ 4 \pi \left< \int_0^{\infty} r^2 \phi {\partial V \over \partial \phi}
 \, dr \right> = 0
\end{equation}
and after integrating by parts the time integral in the first term,
(the boundary term $\phi\dot\phi $ vanishes due to
the integration over a period),
\begin{equation}
-{1\over T} \int_T dt\, 4 \pi \int_0^{\infty} r^2 \dot\phi ^2\, dr + 2 \langle
E_s\rangle
+ 4 \pi \left< \int_0^{\infty} r^2 \phi {\partial V \over \partial \phi}
 \, dr \right> = 0  ~~.
\end{equation}
Identifying the first term as twice the time-averaged kinetic energy,
we arrive at the virial theorem
\begin{equation}
\langle E_k\rangle = \langle E_s\rangle + 2 \pi \left< \int_0^{\infty} r^2 \phi
{\partial V \over \partial \phi} \, dr \right> ~~.
\end{equation}

For the SDWP, in dimensionless variables,
\begin{equation}
\langle E_k\rangle = \langle E_s\rangle +
   2 \pi \left< \int_0^{\infty} \rho^2 \Phi^2 (\Phi^2 - 1) \,
d\rho \right>\, .
\end{equation}

As usual, the virial theorem holds as an equality only for strictly
periodic systems. Numerical simulations of oscillons show, however,
that the basic oscillation is overlaid by a long-wavelength modulation
and other deviations from strict periodicity.
It is hence of interest to analyze
the ``departure from virialization'',
\begin{equation}
{\cal V}(\tau) \equiv \langle E_k\rangle - \langle E_s\rangle -
2 \pi \left< \int_0^{R_s} \rho^2 \Phi^2 (\Phi^2 - 1) \,
d\rho \right>\, ,
\end{equation}
where now
\begin{equation}
E_k  =  2 \pi \int_0^{R_s} \rho^2 \dot\Phi^2 \, d\rho,\;
E_s  =  2 \pi \int_0^{R_s} \rho^2 (\Phi')^2 \, d\rho
\end{equation}
and $R_s \approx 10$ as before is an integration cut-off large enough to
encompass the entire configuration. For a  perfectly virialized configuration,
${\cal V}(t)=0$. In Fig.~12 we show the evolution of ${\cal V}$ for several
Gaussian bubbles. When contrasted with Fig.~6a, it becomes clear that {\it the
longer the lifetime of the oscillon, the better virialized it is}. This
result is made more transparent
by plotting the lifetime as a function of the maximum
value of ${\cal V}$ for several radii, as shown in Fig.~13. Note also the
symmetry about the longest-living oscillon, with $R_0=2.86$.

Using the virial relation and the numerical results, we can obtain a
semi-analytical estimate for the
optimal radius
for an oscillon, that is, the one which is longest-lived.
Although we perform the
calculation for the SDWP, our methods can be easily generalized for any
potential. With the {\it ansatz} for the Gaussian profile given in
Eq.~\ref{ansatz}, the time-averaged oscillon energy, and
the departure from virialization, ${\cal V}$,
are, respectively,
\begin{equation}
{{\langle E\rangle }\over {\pi^{3/2}}}=\left (
{{\sqrt{2}}\over 2}\langle\dot q^2\rangle
+\sqrt{2}\langle q^2\rangle - {{8\sqrt{3}}\over 9}
\langle q^3\rangle +{1\over 2}\langle q^4\rangle
\right )R^3 +{{3\sqrt{2}}\over 2}\langle q^2\rangle R~,
\end{equation}
and
\begin{equation}
{{{\cal V}}\over {\pi^{3/2}}}=\left ({{\sqrt{2}}\over 2}\langle\dot q^2\rangle
+2\langle q\rangle -
{{5\sqrt{2}}\over 2}\langle q^2\rangle+{{16\sqrt{3}}\over 9}\langle q^3\rangle
-\langle q^4\rangle\right )R^3 - {{3\sqrt{2}}\over 2}\langle q^2\rangle R~~.
\end{equation}
Multiplying the expression for the time-averaged energy by $2$ we can eliminate
the cubic and quartic terms in the expression for ${\cal V}$. Using that
for the longest-lived oscillon ${\cal V}\simeq 0$, we obtain a cubic equation
for the optimal radius, $R_{\rm max}$,
\begin{equation}
\left ({{3\sqrt{2}}\over 2}\langle\dot q^2\rangle -{{\sqrt{2}}\over 2}
\langle q^2\rangle
+2\langle q\rangle\right )R^3_{\rm max}+{{3\sqrt{2}}\over 2}
\langle q^2\rangle R_{\rm max} -2\pi^{-3/2}\langle E\rangle \simeq 0~~.
\end{equation}
To proceed, we further assume that $q(\tau)$ is periodic, which is a good
approximation for the longest-lived oscillon. Writing $q(\tau)=
q_0{\rm cos}(\omega \tau )$, with $q_0$ an
amplitude determined numerically (the reader should be
careful to distinguish between this $\omega$ and the one used in the linear
perturbation analysis), the time-averaging can be performed and we finally
obtain,
\begin{equation}
{{\sqrt{2}}\over 4}\left (3\omega^2-1\right )q_0^2
R^3_{\rm max}
+{{3\sqrt{2}}\over 4}q_0^2R_{\rm max}-
2\pi^{-3/2}\langle E\rangle \approx 0~~.
\end{equation}
The roots of this equation are determined once we know the values of the
parameters $\langle E\rangle$, $q_0$, and $\omega$. These can be obtained
numerically using the remarkable independence of oscillons on initial
conditions. We use $\langle E\rangle
\simeq 43$, and $\omega = 2\pi/T\simeq 1.37$.
The maximum
core amplitude, $\Phi_c$, is roughly bounded by
$-0.1\lesssim \Phi_c\lesssim 0.2$,
which
gives for $q_0$ the range $0.45\lesssim
q_0\lesssim 0.6$.
With these parameters, we find that the equation has only one real root,
bounded by $2.90\lesssim
R_{\rm max}\lesssim 3.54$.
This range of values is in excellent
agreement with the observed numerical range for the oscillon radius
(see Fig.~5) providing strong support to our arguments. It also gives the
correct range of initial values for the radius of bubbles which will relax into
the longest-lived oscillons.  Thus,
the oscillon can be interpreted as the attractor field configuration which
minimizes the departure from virialization.

\section{Evolution of Subcritical Bubbles in ADWP: Numerical Results}

It was first noted in Ref.~\cite{PULSONMG} that oscillons will also be
present for nondegenerate potentials. Most of the analytical arguments
above will also apply in this case. In particular, the minimum radius
for subcritical bubbles to evolve into oscillons can also be obtained by the
perturbation analysis presented in Section 4.A. The sufficient conditions
for the existence of oscillons will still be the same, namely, that the initial
energy be above the plateau energy, and that the initial
core amplitude be above the
inflection point of the potential. Of course, the plateau energy will depend
on the degree of asymmetry of the potential.
The important difference is that for
ADWPs, the $O(3)$-symmetric equations of motion admit static solutions
known as bounces \cite{COLEMAN}. These are the well-known critical bubbles
of strong first order phase transitions, which specify the thermal barrier
for the decay of metastable states, $E_{\rm crit}$ \cite{LINDE};
bubbles with radii larger than
critical will grow, converting the metastable phase into the stable phase
with lower free-energy density. Thus, when discussing oscillons in the
context of ADWPs, we must make sure that the initial configurations have radii
smaller than the critical bubble radius, $R_{\rm crit}$, as well as energies
smaller than the decay barrier. The initial bubble energy is
bounded by the plateau
energy from below and the decay barrier from above.

In order to see the effects of the asymmetry on the properties of the
oscillons, we start by showing the results for the degenerate case,
obtained by setting $\alpha = 3/\sqrt{2}$ in Eq.~\ref{eqmotsp}. Recall that
in this case the minima are at $\Phi_0=0$ and $\Phi_+=\sqrt{2}$. In Fig.~14
we show the lifetime of oscillons as a function of initial radius
for several core amplitudes. Note that
the lifetimes are larger than for the SDWP (Fig.~6a).
This is simply due to the fact
that for $\alpha=3/\sqrt{2}$ the ADWP is shallower and narrower than the
SDWP, softening the surface energy of the initial bubbles.
In Fig.~15a we show the lifetimes vs.~radii of initial Gaussian bubbles
leading to long-lived oscillons
for different values of $\alpha$.
For reference we also show the values of
the critical radii. The perturbative analysis of Section 4.A can easily
be adapted to this ADWP case, yielding an expression for the frequencies
$\omega^2({\bar q},R)$ of small fluctuations (analogous to Eq.~\ref{omegaeq}),
\begin{equation}
\omega^2({\bar q},R) \equiv
{3\sqrt{2}\over 4}{\bar q}^2- {4\sqrt{6}\alpha\over 9}{\bar q}
+\left (1 +{3\over {R^2}}\right )~~.
\end{equation}
The minimum of this surface (for fixed $\alpha$) is once again independent of
$R$ with ${\bar q}_{\rm min} \simeq 0.51 \alpha$, hence oscillons are
possible only for $R_{0} > R_{\rm min} \simeq (3/(0.28\alpha^2 -1))^{1/2}$.
With the values $\alpha = 3/\sqrt{2},~2.16~
{\rm and}~ 2.23$ we then obtain $R_0 > 3.39,~3.12~{\rm and}~2.76$
respectively, results which compare favorably with the numerical simulation
values of $R_0 > 3.2,~3.1~{\rm and}~2.9$, respectively.
In Fig.~15b we show lifetime vs.~initial bubble
energy for different values of $\alpha$. For reference we give the
values of the plateau energy and of the decay barrier.
Note that as the asymmetry is increased, the lifetimes
of the oscillons also increase, almost by a factor of two between
the nearly degenerate $\alpha=2.16$ and the more asymmetric
$\alpha=2.23$, while the ratio between
the critical bubble radius, $R_{\rm crit}$, and the longest-lived oscillon,
$R_{\rm max}$, varies from $R_{\rm crit}/R_{\rm max}\simeq 5$ for
$\alpha=2.16$ to $R_{\rm crit}/R_{\rm max}\simeq 2$ for $\alpha =2.23$.
As the asymmetry is increased, the oscillons approach more and more the
critical bubble, explaining their increased longevity.

\section{Oscillons in Action: Possible Applications}

In this Section we will present a few
situations in which we expect oscillons to be relevant. As we will argue below,
their remarkable
longevity
makes them specially interesting in the context of
phase transitions; if thermally nucleated, their presence can affect the
dynamics of the transition in several ways.
It is not our intention here
to give a detailed treatment  of the r\^ole of oscillons on the
dynamics of phase transitions, but simply to stress the interesting physics
that can emerge due to these configurations.

As we have seen, a typical range of
lifetimes is between $t_l=10^3 - 10^4~m^{-1}$
in units of the mass $m$ introduced
in Eqs.~\ref{potsdwp} and \ref{potadwp}. This is much longer than that of the
solution to the spherically symmetric linear Klein-Gordon equation $\sim
5 m^{-1}$. The expansion rate of the Universe in
a radiation-dominated regime can be written in terms of the background
temperature $T$ as $H^2\propto T^4/m^2_{\rm Pl}$, where $H$ is Hubble's
parameter. Thus, the expansion time-scale is $t_U\sim H^{-1}\propto
(m_{\rm Pl}/T)T^{-1}$. Typically, the symmetry breaking temperature $T_c$
can be
written in terms of the mass scale $m$ of the theory as
$T_c\simeq m/\sqrt{\lambda}$. Thus, the expansion time-scale at $T_c$ is
$t_U\sim \lambda (m_{\rm Pl}/m) m^{-1}$. The ratio between the oscillon
lifetime
(taking $t_l=10^4~m^{-1}$) and the expansion time-scale is then,
$t_l/t_U\sim \lambda^{-1} 10^4 (m/m_{\rm Pl})$.
{}From this we
see that for masses of order the GUT scale the lifetime of the oscillons is
comparable (or larger, for weak coupling!)
to the age of the Universe at that scale, an intriguing possibility.
In such a scenario these unstable field configurations could have a dramatic
effect on the dynamics of any phase transition.
For example, during a first order
phase transition those subcritical bubbles which go on to
form oscillons could last long
enough to become critical bubbles as the Universe cooled. In
this way we would have a method of completing the transition quicker.
Another possibility is that oscillons act as seeds or
nucleation sites for the critical
bubbles. The combination of these two effects will increase
the production rate of critical bubbles, a
feature which may well have useful consequences for the old inflationary
universe scenario. That model failed partly
because the production rate of
critical sized
bubbles and larger could not keep pace with the exponential
expansion of the
Universe. If the bubble nucleation rate were increased, then this problem may
well be overcome.

For oscillons to be relevant cosmologically, not only must they survive for
long enough, they must be thermally produced in large enough numbers
since they are
unstable and eventually decay. A naive estimate of this
rate is that the number density of oscillons of size $R$ produced
at temperature $T$ due to thermal fluctuations is
$$n(R,T) \sim T^3 e^{-F(R)/T}~,$$ where $F(R)$ is the free-energy of the
configuration of radius $R$ and is given by $F(R) = E_s + E_v$ in
Eq.~\ref{enterms}.
Comparing $F(R_{osc})$ with $F(R_{cr})$ gives an indication
of the fraction of bubbles which are oscillons as opposed to critical at
any given temperature $T$. In fact we can see quite easily that although
the oscillons are unstable they are
produced in much greater abundance, as their free-energy barrier is typically
smaller than that for critical bubbles.
To be sure of this we
require that their thermal nucleation rate be considerably larger than the
expansion rate of the Universe, {\it i.e.}
$\Gamma_{\rm th}(R,T)/H \gg 1$. Since
$H \propto T^2 / m_{\rm Pl}$, it becomes a straightforward comparison.
For a GUT scale transition, say with $T_c \sim 10^{15}$ GeV,
this condition implies $F_{\rm osc}/T_c < 10$,
which is not difficult to satisfy for sufficiently weak transitions, such as
the SU(5) Coleman-Weinberg model, as shown in Ref. \cite{GKW}.
More speculatively, even for lower energies, oscillons may be potentially
relevant. Although in these cases
their lifetime becomes small compared to the expansion
time-scale, if they are produced in large enough numbers these sub-critical
bubbles could still affect the dynamics of the transition. The presence of
oscillons will
substantially increase the equilibrium number-density of sub-critical
bubbles of the broken phase; as they last longer, their depletion rate by
shrinking is decreased, and hence the net volume in the broken phase
increases. (In Ref. \cite{GG}, it was assumed that all subcritical bubbles were
roughly of a correlation volume and disappeared in a time $\sim \xi(T)$, where
$\xi(T)$ is the correlation length.)
They effectively make the transition weaker than what one would predict
from the effective potential. Also, the collision of oscillons with expanding
critical bubbles will possibly cause instabilities on the bubble wall, implying
that the assumption of spherical evolution of the walls may be incorrect. For
the case of the electroweak transition, Gleiser and Kolb \cite{GK} have shown
that the condition on subcritical bubbles can be written as $F(R)/T < 34$.
This in turn imposes a constraint on the mass of the associated Higgs, which
for the one-loop potential
turns out to be $m_{\rm Higgs} \ge 88 $GeV (see \cite{CGM} for details).
It seems to be the case that oscillons will have an important effect on the
dynamics of sufficiently weak first order phase transitions.

In this paper we have been investigating the existence of oscillons for both
first and second order phase transitions. A number of issues arise common to
both cases which require further study. The first concerns the coupling
of oscillons to other forms of matter, whether they be other scalar fields,
gauge fields or fermions. We have regarded the oscillons as emerging
from an effective theory in which the fields to which it is coupled have
been integrated out (a procedure we would hope is valid for low
enough energies), leaving an effective potential for the scalar field.
Ideally we would like to consider the full theory and solve for all the fields
without integrating out the massive ones. It could be that one of
these fields leads to an instability in the $\Phi$ field which causes the
oscillon to decay faster than we have estimated. On the other hand, coupling
the oscillon to a charged field may enhance its lifetime,
as in the case of nontopological solitons
\cite{NTSs}.  A second issue concerns the coupling of oscillons to hot
plasmas, as would be the case during thermal phase transitions. The plasma
would act both as a viscous medium and as an enhancer of fluctuations,
presumably affecting the lifetime of the oscillons. We are currently
investigating both issues.

The discussion in this Section has concentrated on early Universe aspects of
oscillons. Since they are field theories we should expect them to be seen
at laboratory energies as well. This may not be so easy to do in practice,
but there
are many examples of phase transitions in liquid crystals and Ising-like
systems which
produce nonlinear field theory objects such as topological defects
\cite{NEIL}. Also,
solutions to the nonlinear Schr\"odinger equation have been known to be
of importance in several contexts, including the propagation of
information in optical fibres \cite{FIBRES}.
It is reasonable to expect that oscillons will
be present in the
non-relativistic limit, thus being possible solutions to the time-dependent
nonlinear Schr\"odinger equation as well. What
would be required for oscillons is a distinct signature. It
could be that as the energy of the system reaches its plateau during the
oscillon stage, the material has a particular refractive
index and thus could be detected in scattering experiments.

\section{Conclusions and Outlook}

In this paper we have presented the results of a detailed investigation of
the properties of oscillon configurations and explained, where possible,
the physics behind their interesting dynamics.
The fact that they exist in both
first and second order phase transitions makes them of particular interest.
They are localized, non-singular, time-dependent, spherically-symmetric
solutions of nonlinear scalar field theories, which are unstable but extremely
long-lived, with lifetimes of order $10^3-10^4~m^{-1}$, where $m$ is the
mass of the scalar field. They naturally appear during the collapse of
spherically symmetric field configurations.
We have obtained the conditions required
for their existence, namely that
the initial energy needs to be above a plateau energy and the initial
amplitude of the field needs to be above the inflection point on the
potential in order to probe the nonlinearities of the theory (but does
not need to be at the true minimum of the potential).

Of the many
intriguing aspects of these configurations, some that stand out include the
fact that they exist only
for a given range of initial radii and core amplitudes.
The lower value of the radii can be explained by perturbation theory. It
corresponds to the minimum radius beyond which the field probes the
nonlinearity
of the potential. Explaining the upper bound for the initial radius of the
field
profile is not so straightforward and we are currently investigating this. It
could well be that since larger bubbles have larger initial energies,
during their collapse higher nonspherical modes are excited, triggering the
rapid growth of instabilities responsible for the bubble's collapse before it
can settle into the oscillon stage.
Another remarkable feature is that the plateau energy of the oscillon is
practically
independent of the initial radius. We have interpreted this fact by
showing that the oscillon can be thought of as the attractor field
configuration which minimizes the departure from virialization.

There is much that remains to be investigated. One concern is that we only
investigated stability to radial perturbations. We really need to investigate
how nonspherical perturbations affect the spherically symmetric solutions. One
possibility is that they will tend to make the oscillons collapse into a
pancake configuration, and hence decay more quickly than in the spherical case,
although we believe this will only be the case for bubbles with large initial
radii.
We may also think of higher nonspherical modes as excited states
of the ``ground-state'' $\ell=0$ resonance studied here. It is thus possible
that
oscillons may appear in higher energy configurations, which may decay either
to the ground-state oscillon or just into scalar radiation.
Finally, a more detailed study of the coupling of
these objects to other matter fields and hot plasmas is required in order to
investigate how they affect the dynamics of phase transitions
and how their own decay is affected by these couplings. It is clear though
that they are of interest cosmologically. We are currently analyzing
the consequences of oscillons if they were to be
formed at the electroweak scale \cite{CGM}.

\acknowledgments

We thank Tanmay Vachaspati for stimulating discussion during the
early stages of this work.
This collaboration was partially supported by a NATO Collaborative Research
Grant `Cosmological Phase Transitions' no. SA-5-2-05 (CRG 930904)
1082/93/JARC-501.
EJC was supported by PPARC and would like to
thank the Isaac Newton Institute where part of this work
was completed for financial support and their kind hospitality.
MG thanks the School of Mathematical and Physical Sciences of the University
of Sussex
and the Nasa/Fermilab Astrophysics
Center for their kind hospitality,
where part of this work was completed. At Fermilab MG was partially supported
by NASA Grant NAG 5-2788 and DOE.
MG was partially supported by the National Science Foundation through a
Presidential
Faculty Fellows Award no. PHY-9453431, and by Grant no. PHYS-9204726. He also
acknowledges support from NASA Grant NAGW-4270.
HRM thanks Dartmouth College for a Graduate Fellowship.

\listoffigures

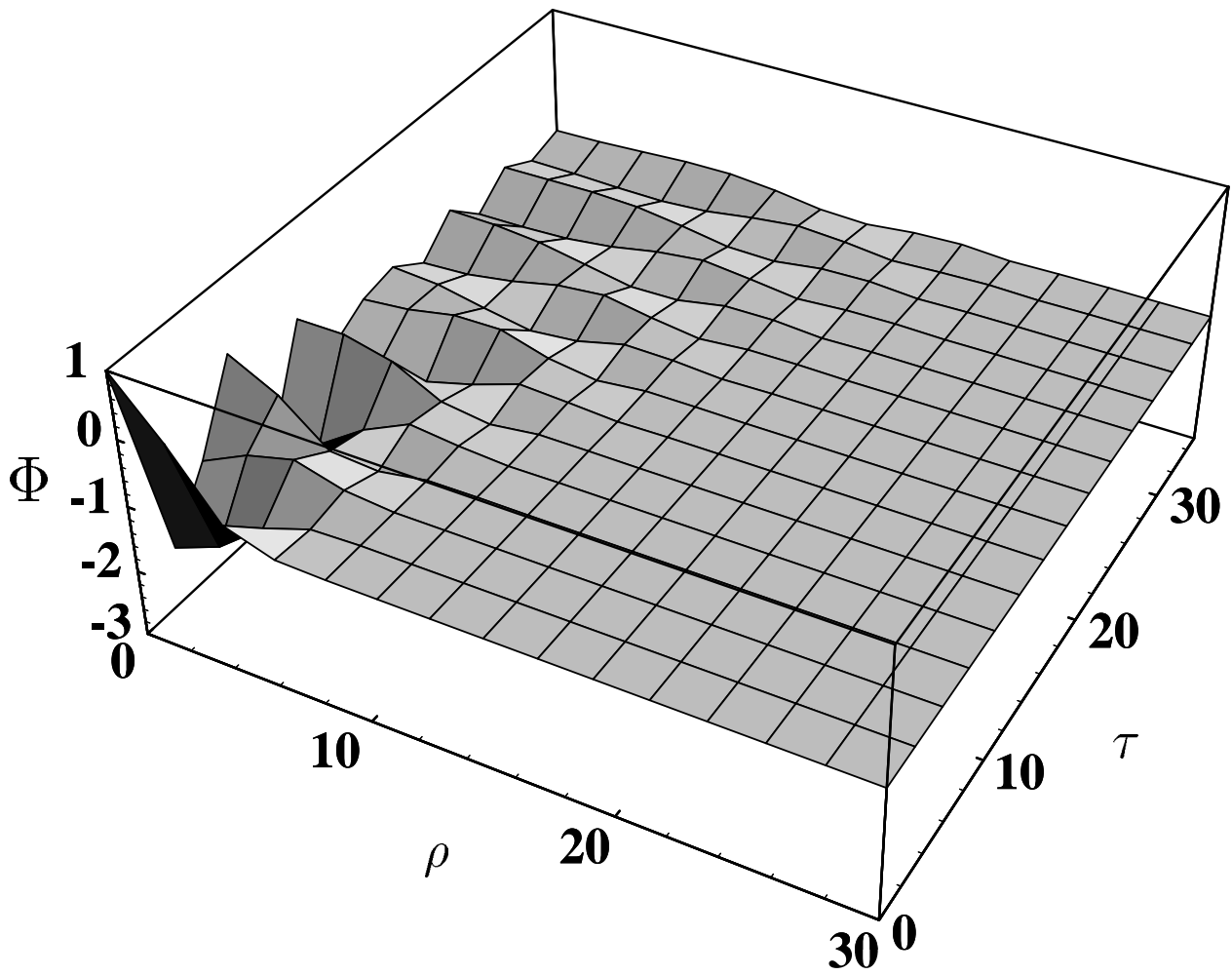
\noindent Figure 1. Solution $\Phi(\rho,\tau)$ of linear Klein-Gordon equation
($R_0
= 3$).

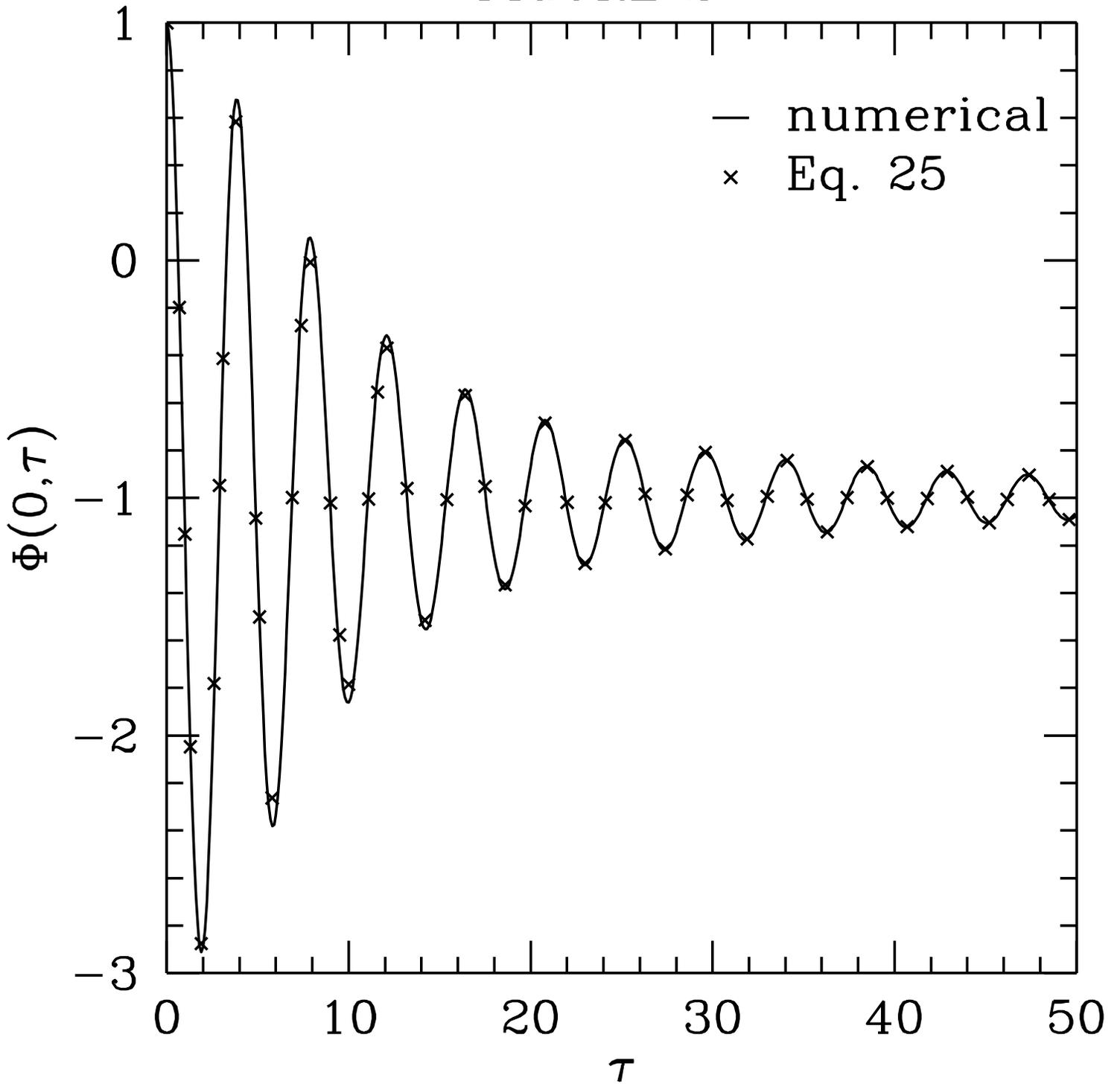
\noindent Figure 2. Comparison between analytical and numerical results
                    for linear solution $\Phi(0,\tau)$ at core ($R_0 = 3$).

\noindent Figure 3a, b. Energy vs.~time for Gaussian and tanh bubbles in the
SDWP.

\noindent Figure 4a. Time evolution of bubble's core during oscillon stage
                    ($R_0 = 2.7$).

\noindent Figure 4b. Snapshots of oscillon with initial radius $R_0 = 2.5$.
                    The snapshots are $\Delta\tau = 0.2$ apart. The dotted
lines are
                    snapshots with $\dot\Phi > 0$ for $\tau > 504$.

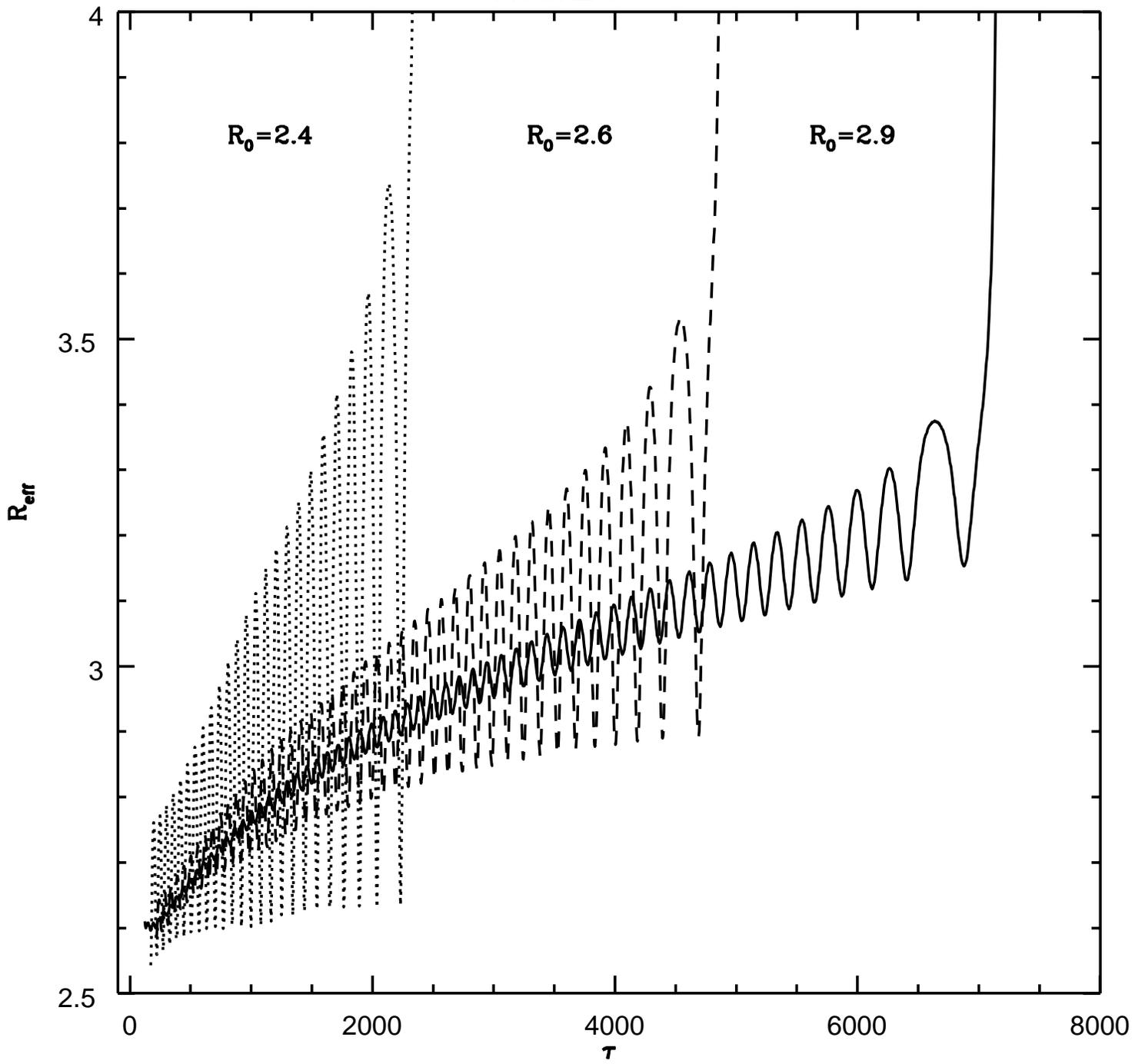
\noindent Figure 5. Evolution of averaged oscillon radius for several
                    initial Gaussian bubbles.

\noindent Figure 6a. Oscillon lifetime vs.~initial radius of Gaussian and
                     tanh bubbles (SDWP).

\noindent Figure 6b. Oscillon lifetime vs.~initial bubble energy.

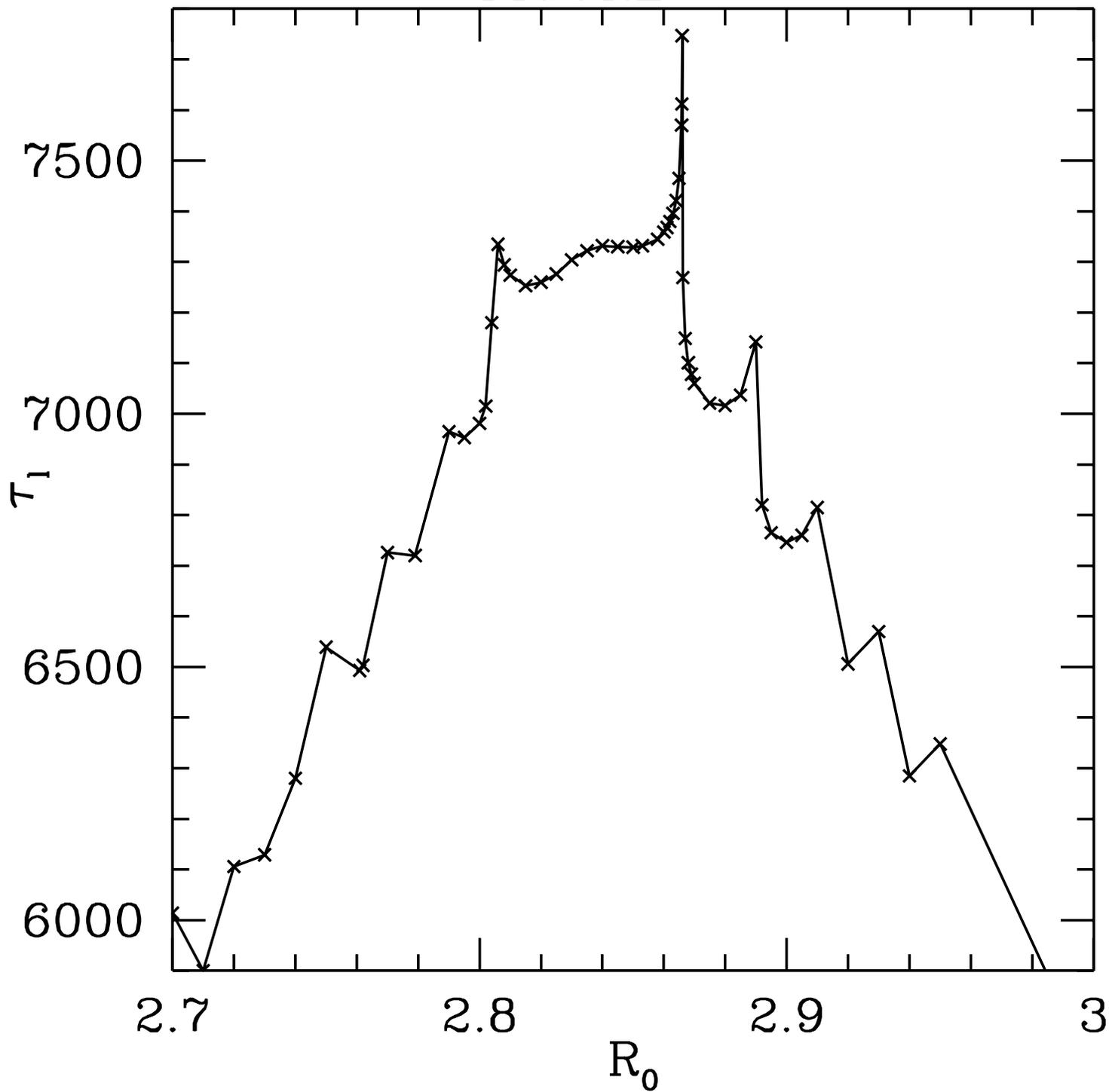
\noindent Figure 7. Detail of oscillons' lifetime around the peak.

\noindent Figure 8a, b. Oscillon lifetime vs.~initial Gaussian radius
                    and vs.~initial energy, respectively,
                     for different initial core amplitudes.

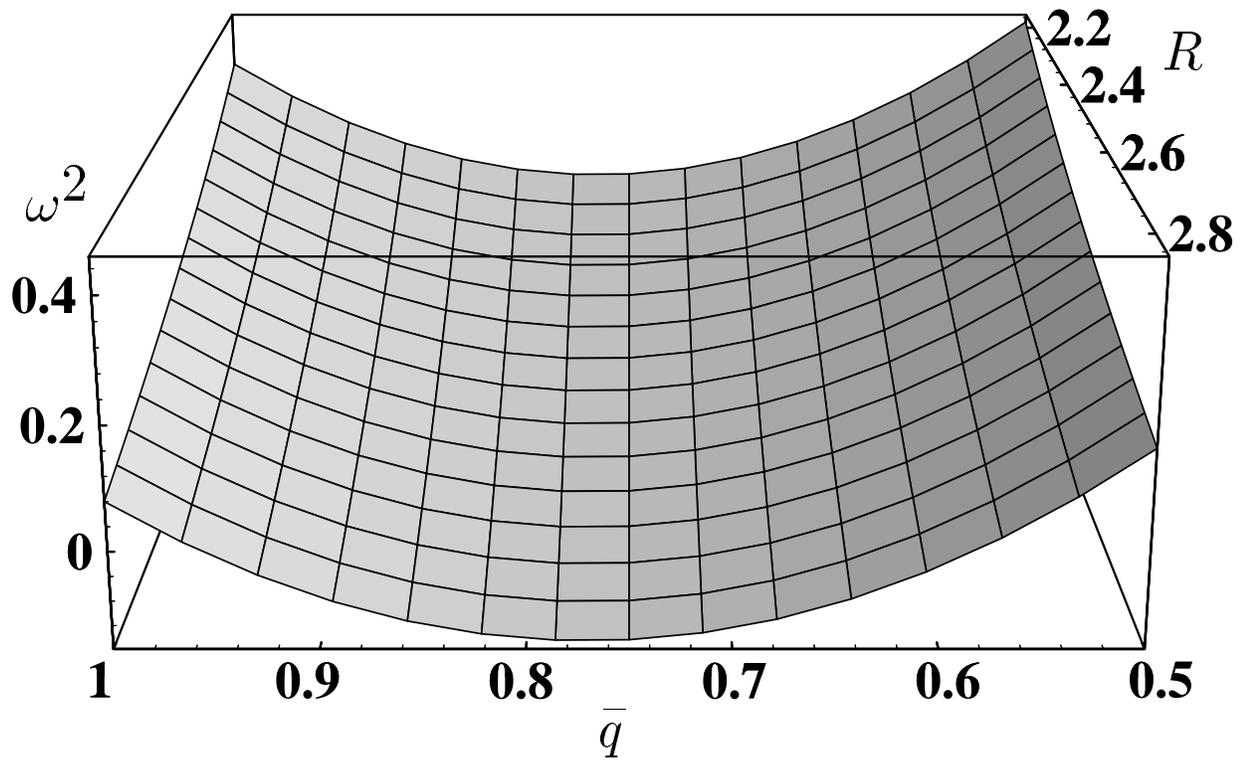
\noindent Figure 9. The frequency surface $\omega^2({\bar q}, R)$.

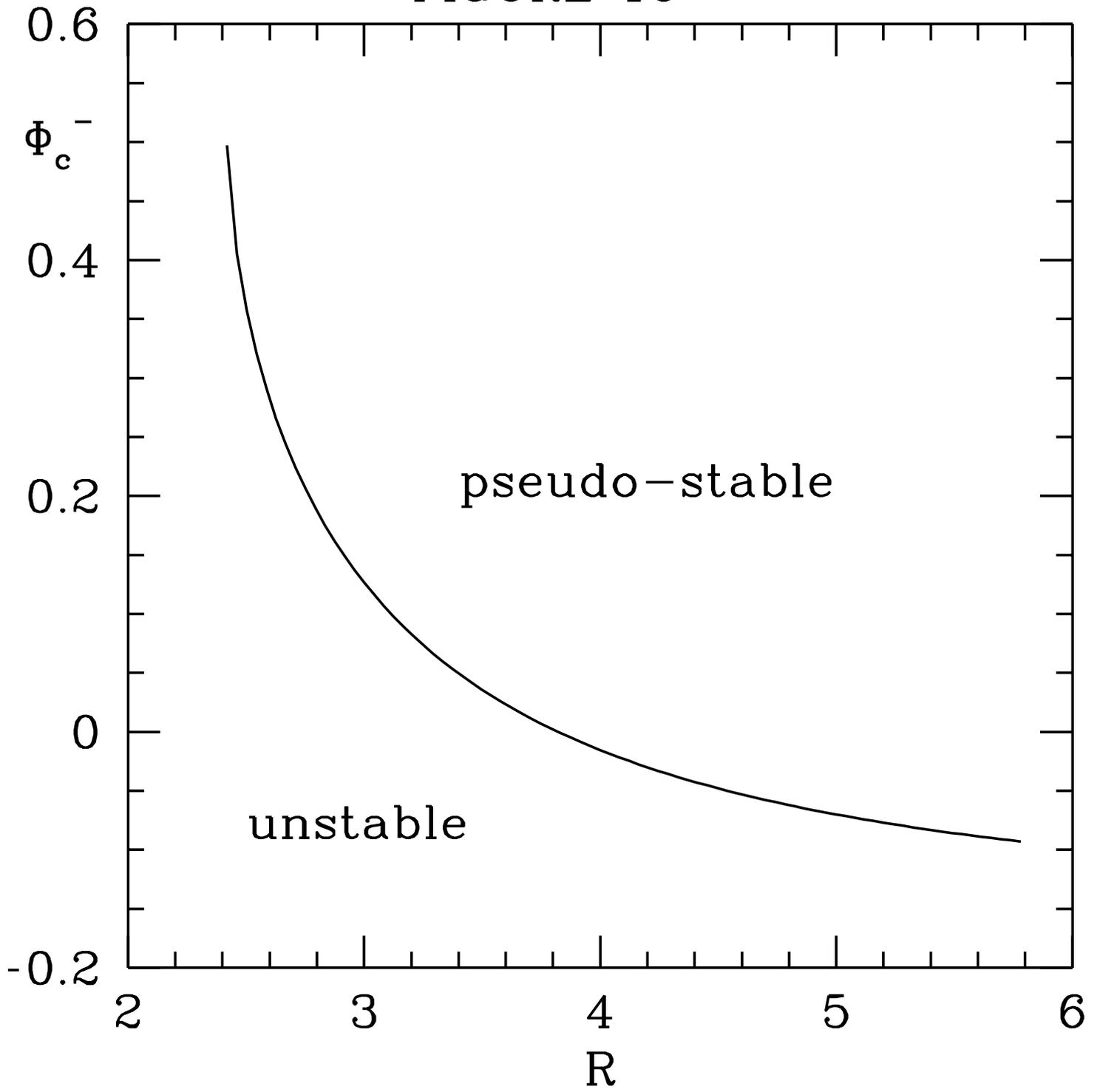
\noindent Figure 10. Minimum amplitude $\Phi_c^-$ for nonlinear growth
vs.~radius.

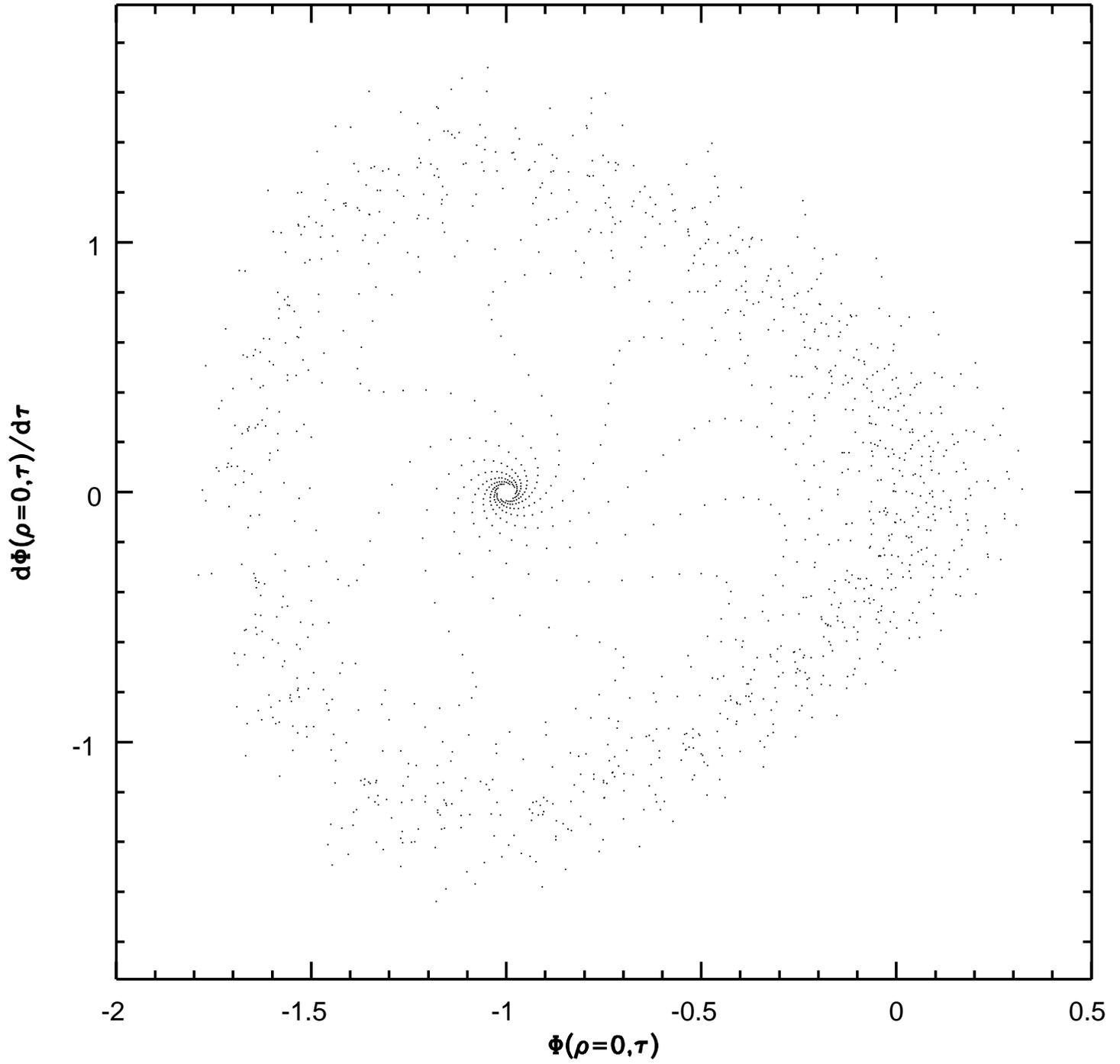
\noindent Figure 11. Phase-space portrait of an oscillon for $R_0=3.0$.
                    The sampling occured every $\Delta\tau = 1.0$.

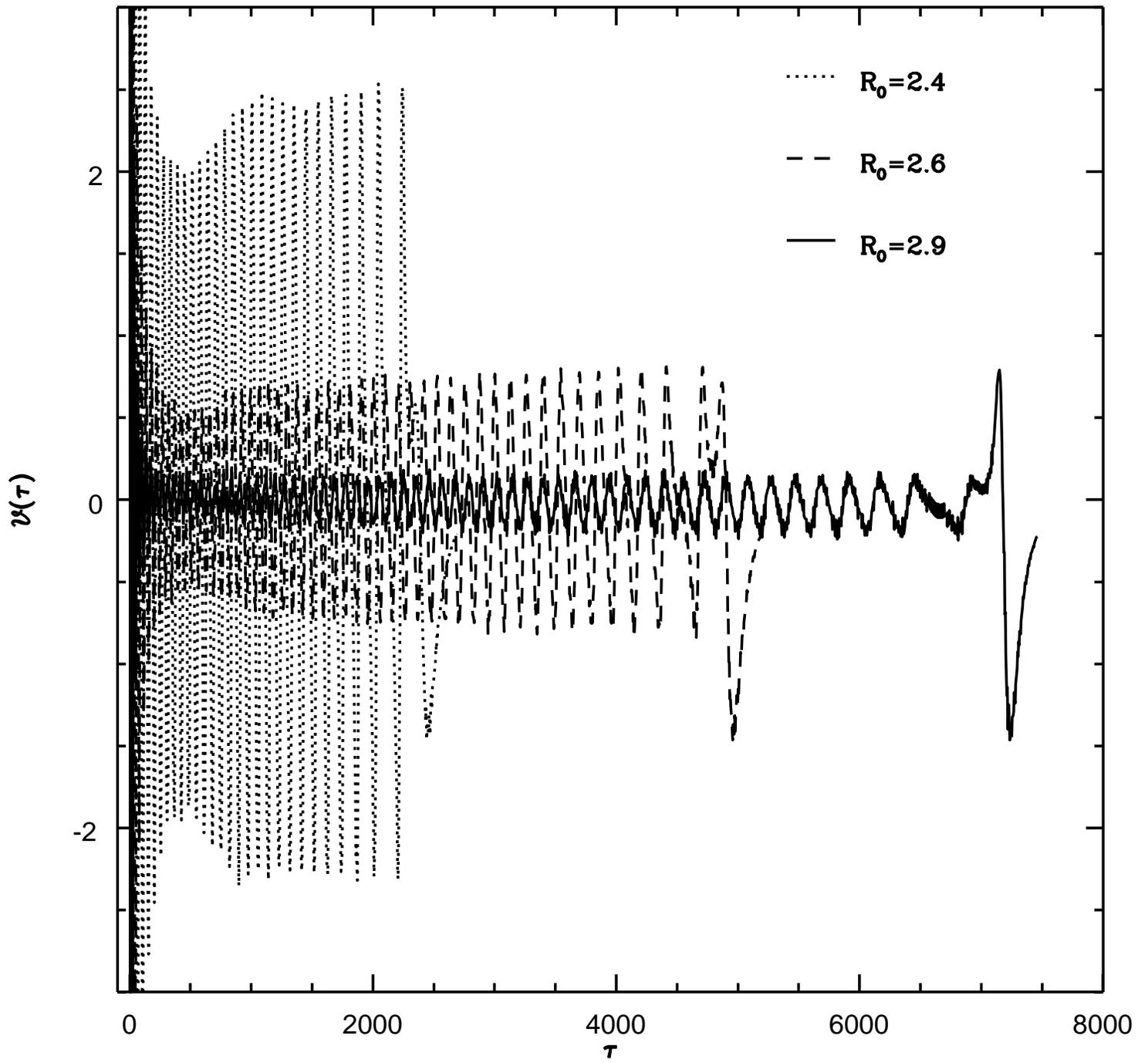
\noindent Figure 12. Departure from virialization vs.~time for several
                     initial bubble radii.

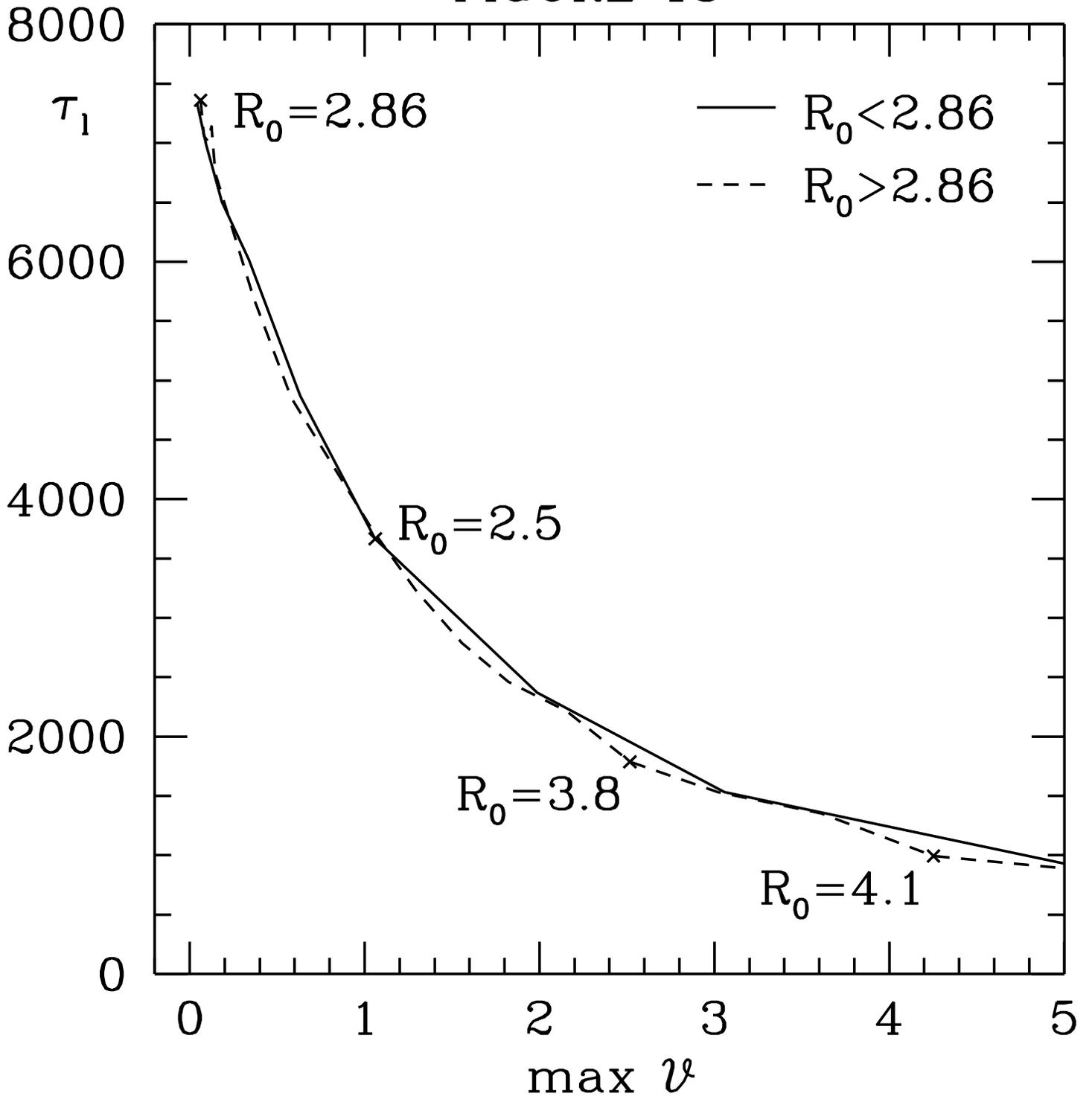
\noindent Figure 13. Oscillon lifetime vs.~maximum value for departure from
                     virialization for several initial radii.

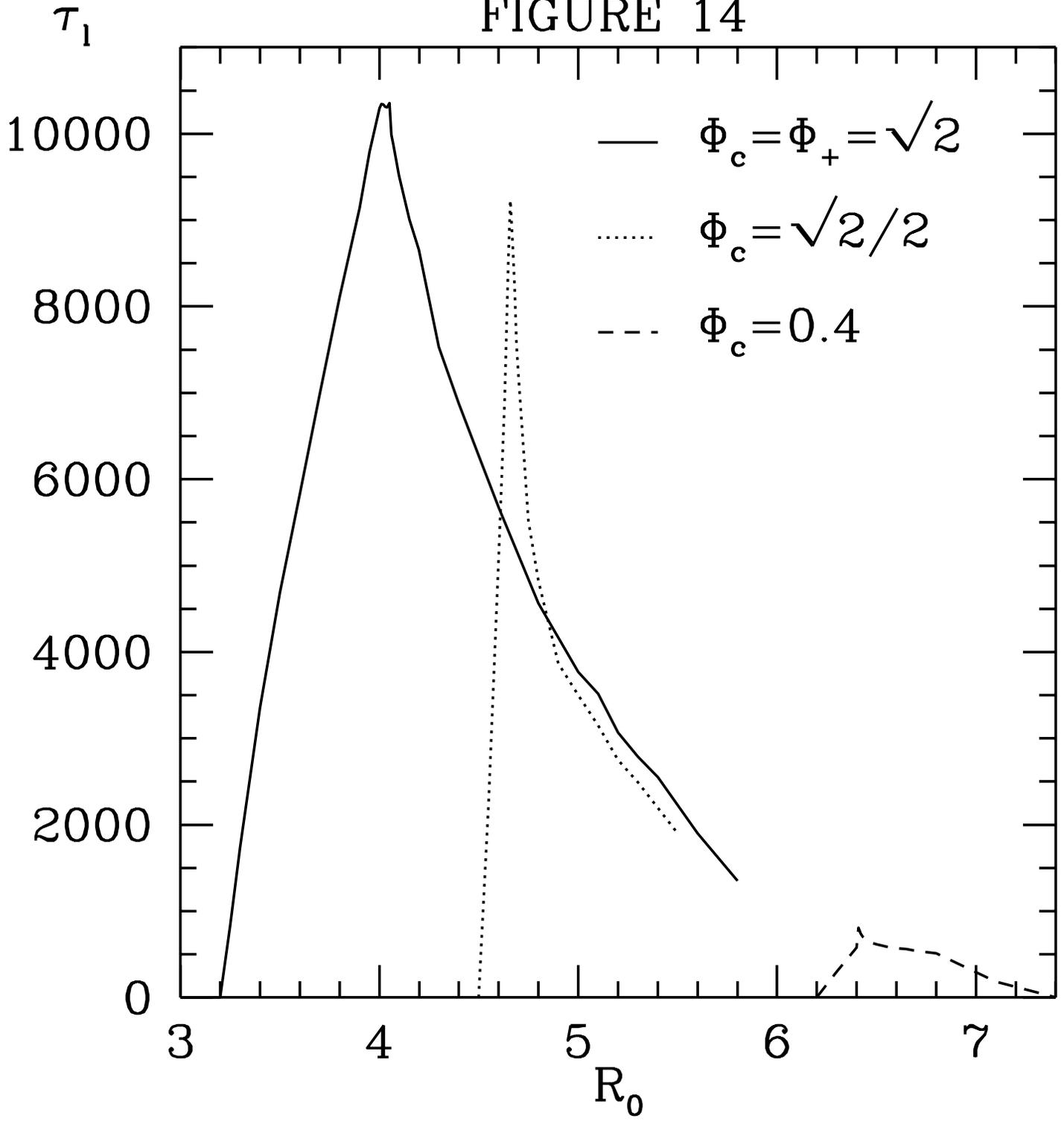
\noindent Figure 14. Oscillon lifetime vs.~radius for several initial
                     core values with $\alpha =3/\sqrt{2}$.

\noindent Figure 15a. Oscillon lifetime vs.~initial radius for several
                      values of $\alpha$. The respective values of the critical
                      bubble radius are also shown.

\noindent Figure 15b. Oscillon lifetime vs.~initial energy for several
                      values of $\alpha$. The respective values of the
                      oscillon's plateau energy and of the decay energy barrier
                      are also shown.

\listoftables

\noindent Table 1. Comparison between analytical value for core amplitude
                   appropriate for oscillon stability (Eq.~\ref{minamp})
                   and numerical value
                   of core's amplitude at last oscillation for several
                   oscillons. The value of the radius used in analytical
                   formula was obtained numerically (see Fig.~5).

\vfill\eject
\pagestyle{empty}

{\Huge TABLE 1}
\bigskip\bigskip
\bigskip\bigskip

{\large
\begin{tabular}
{|@{~}c@{~}|@{~}l@{~--~}l@{~}|@{~}l@{~--~}l@{~}|@{~}l@{~}|}
\hline
$R_0$ & \multicolumn{2}{c}{Range of $R_{\rm eff}$}&
\multicolumn{2}{c}{Range of $\Phi_c^-$} &
{$\Phi_{c,{\rm ~num}}$}\\ \hline
2.4 & 2.635 & 3.737 & 0.00836 & 0.2614 & 0.2705 \\
2.5 & 2.781 & 3.563 & 0.0278  & 0.1946 & 0.1626 \\
2.6 & 2.893 & 3.532 & 0.0316  & 0.1564 & 0.09160 \\
2.7 & 3.010 & 3.458 & 0.0413  & 0.1242 & 0.02574 \\
2.8 & 3.159 & 3.298 & 0.0656  & 0.09091& 0.05090 \\
2.9 & 3.120 & 3.333 & 0.0599  & 0.09892& 0.03163 \\
3.0 & 2.991 & 3.403 & 0.0492  & 0.1290 & 0.03631 \\
3.1 & 2.904 & 3.439 & 0.0440  & 0.1531 & 0.08632 \\
3.3 & 2.778 & 3.575 & 0.0263  & 0.1957 & 0.1640 \\
3.5 & 2.680 & 3.575 & 0.0263  & 0.2380 & 0.2368 \\
3.7 & 2.589 & 3.848 & 0.0237  & 0.2890 & 0.3113 \\
\hline
\end{tabular}
}

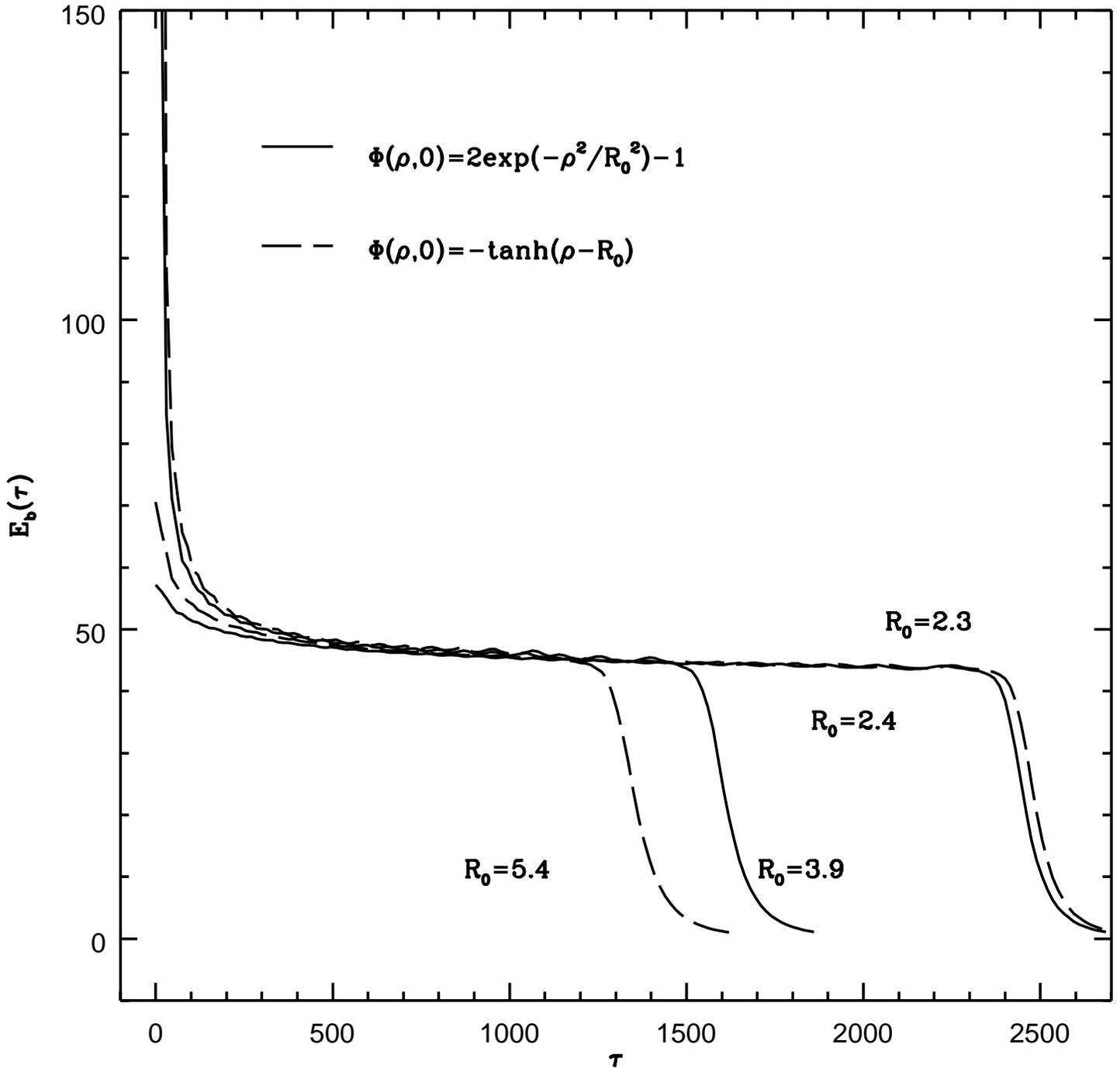
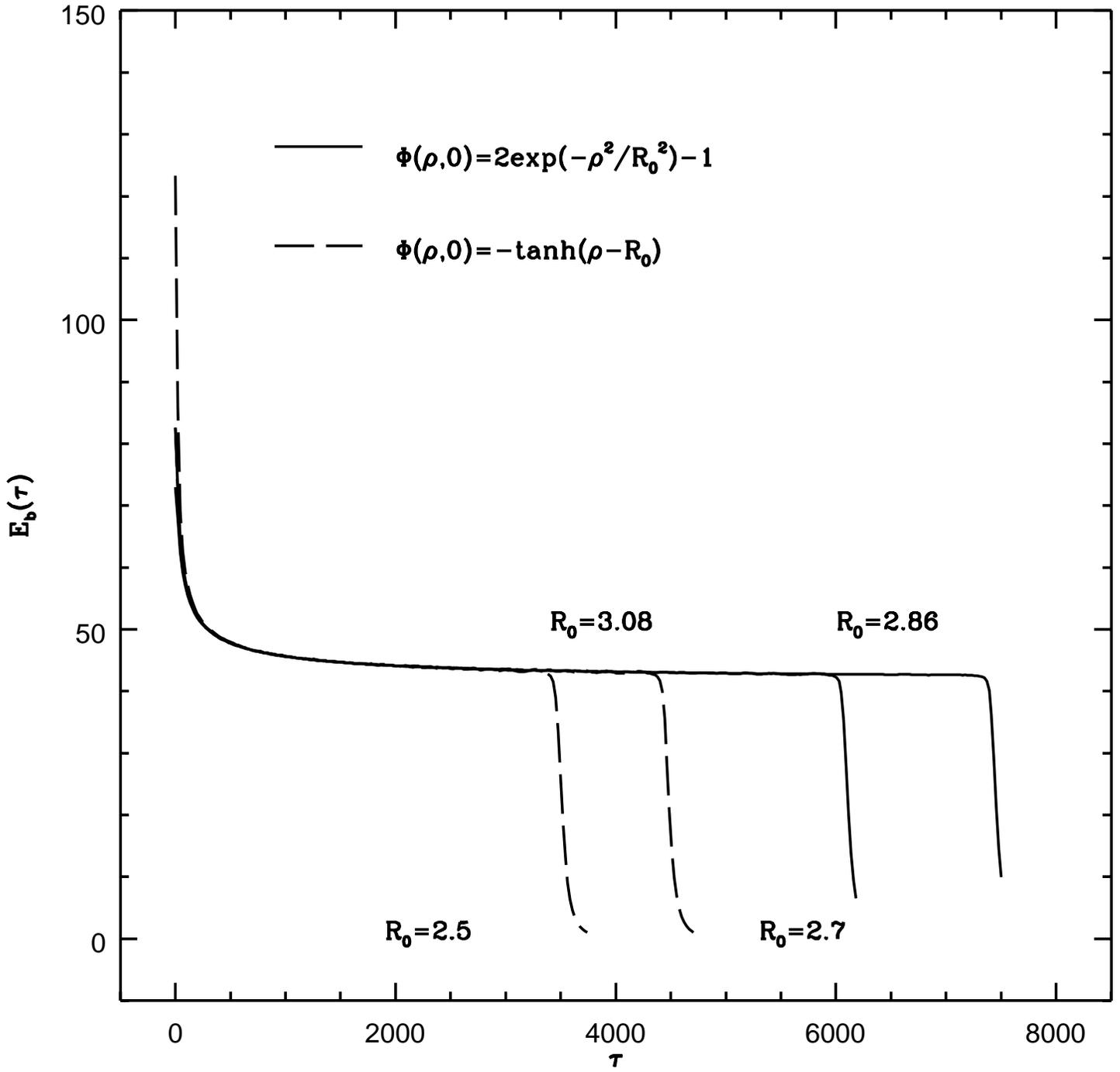
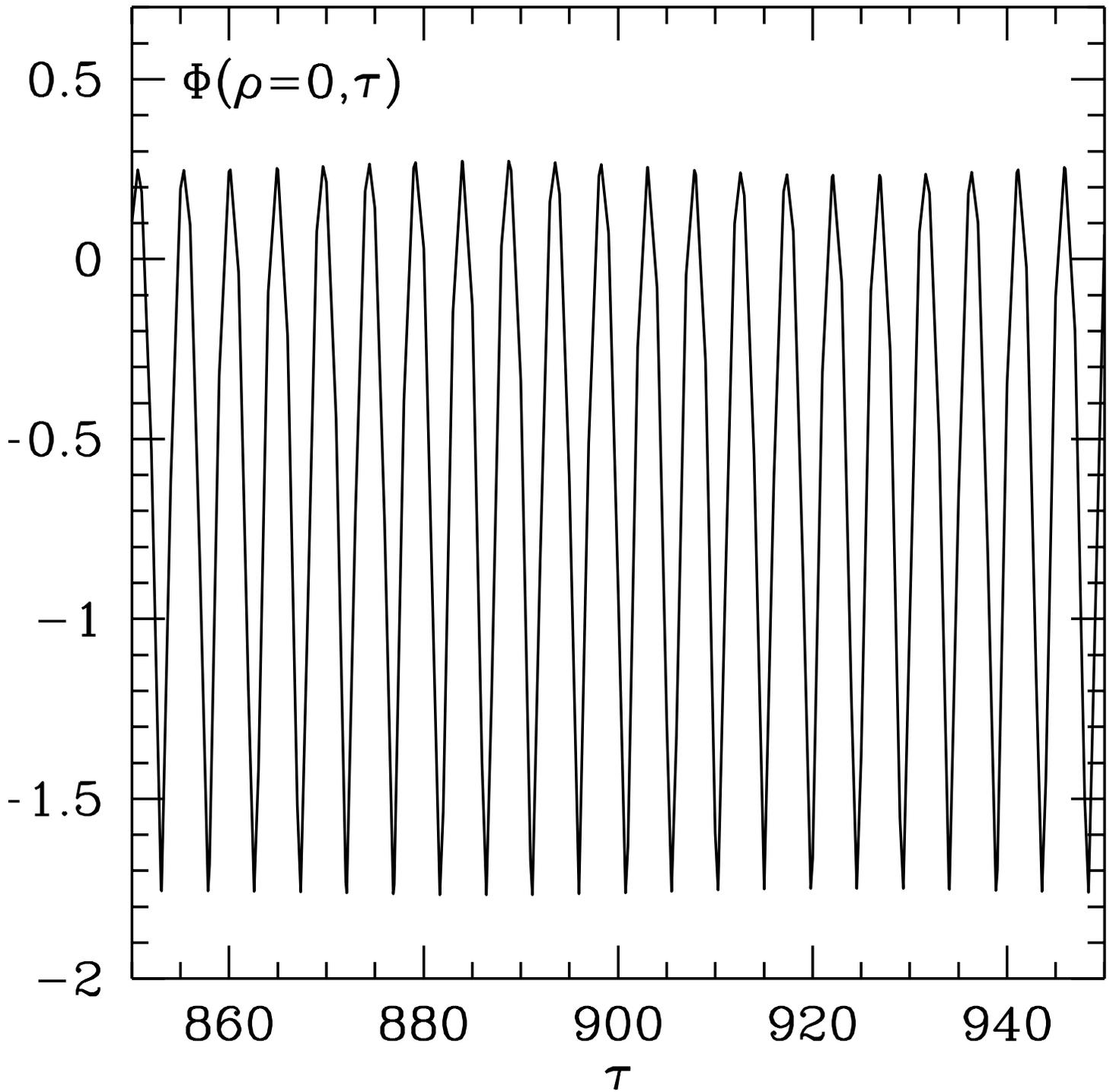
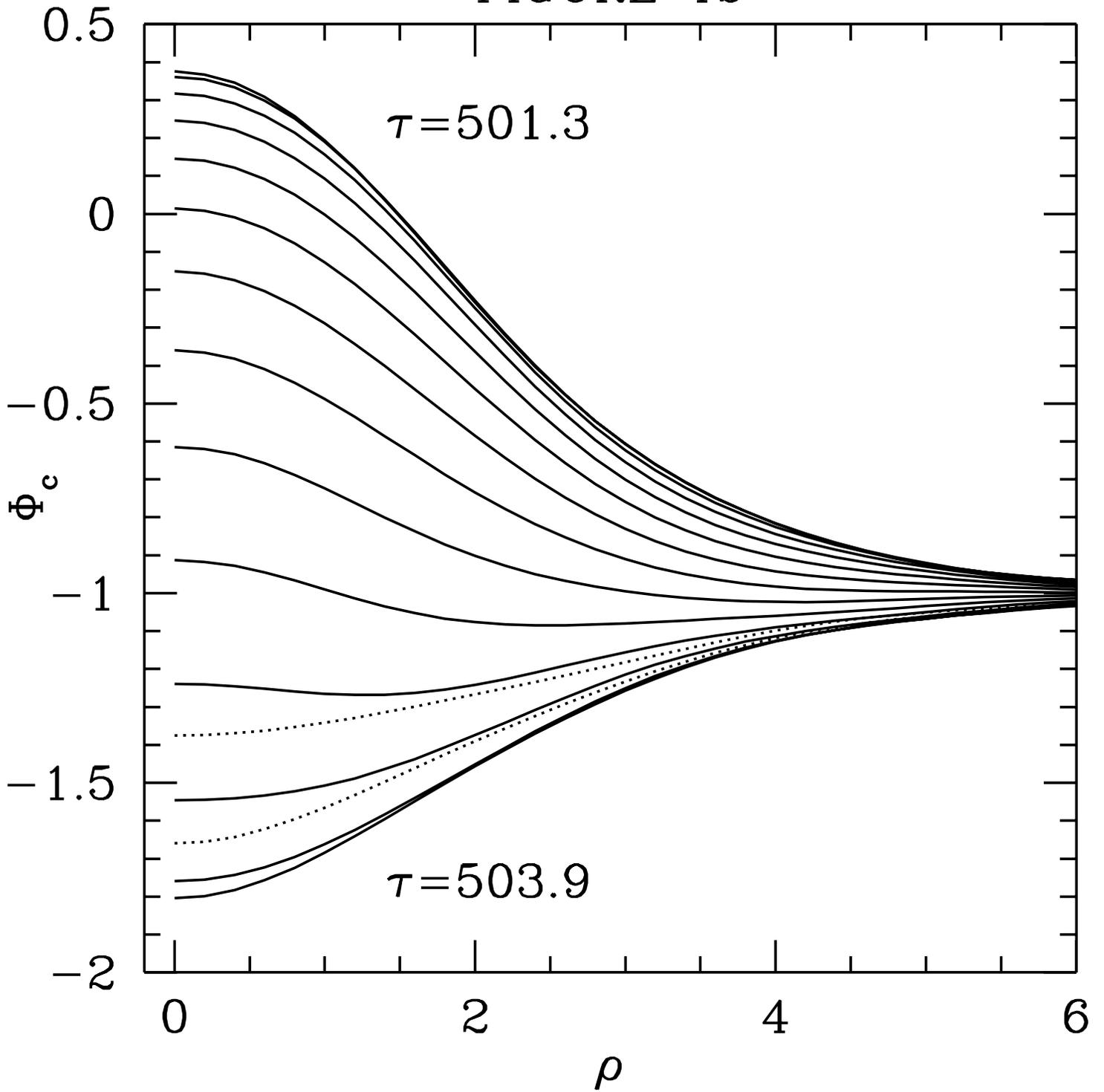
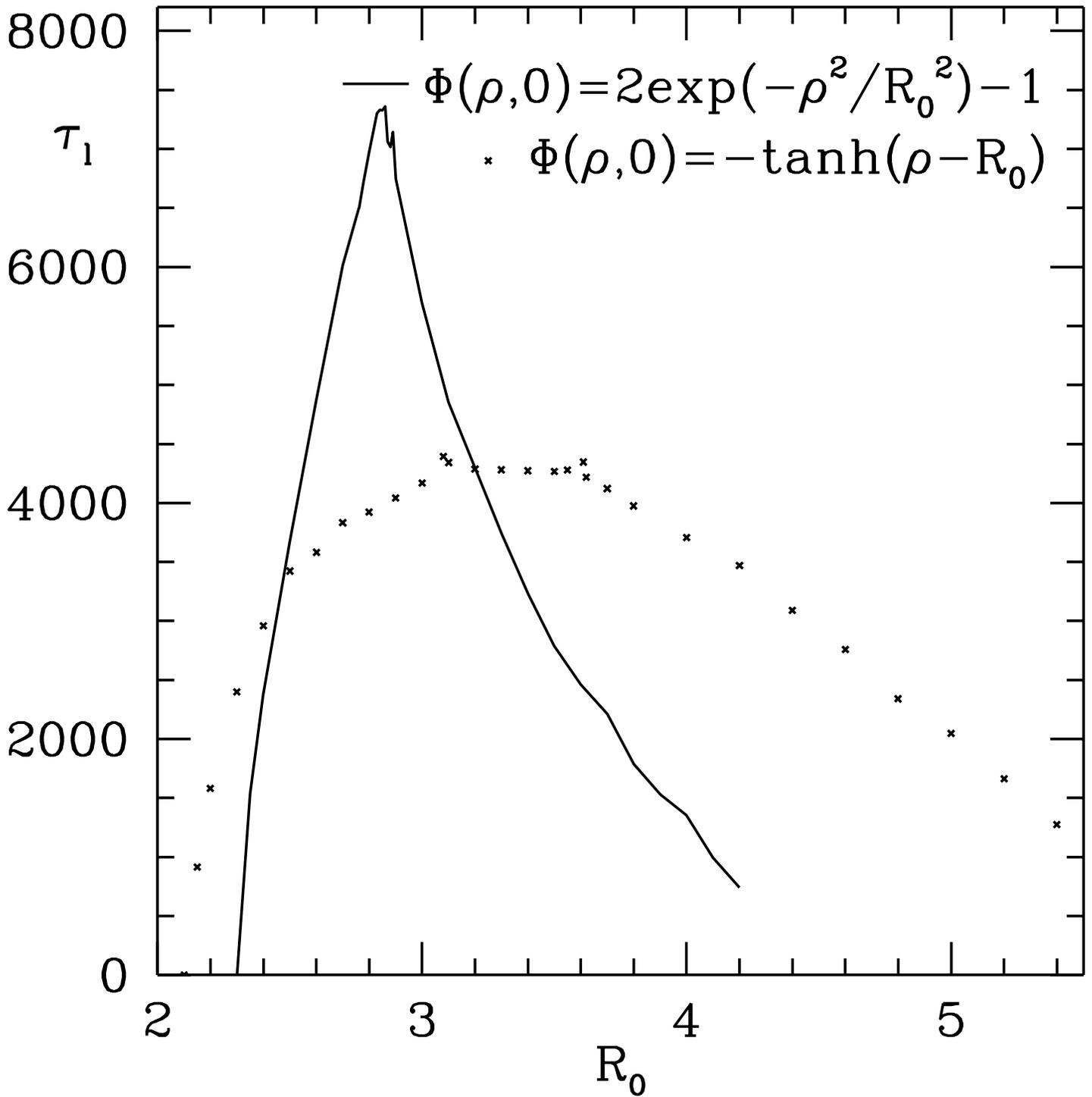
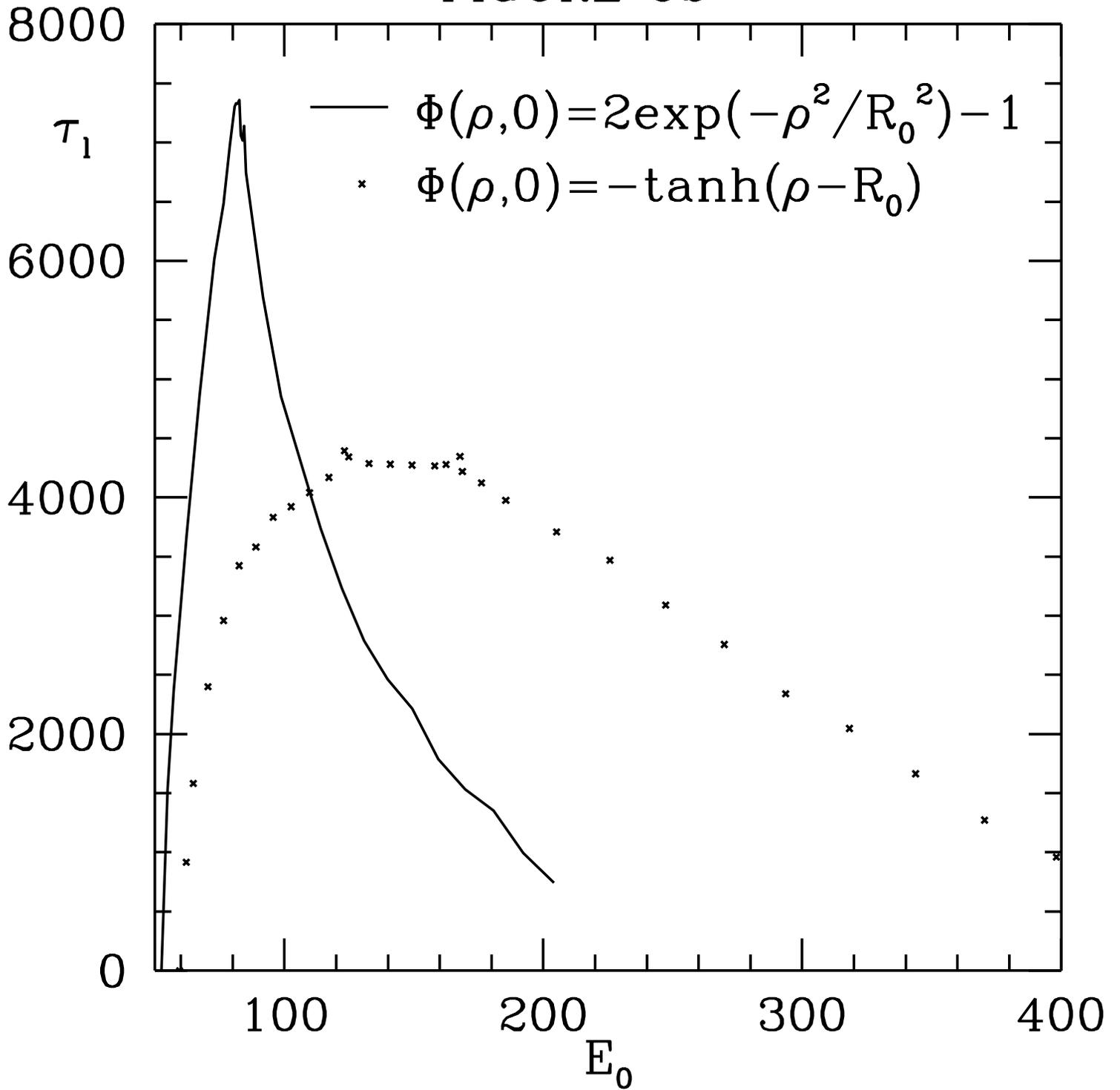
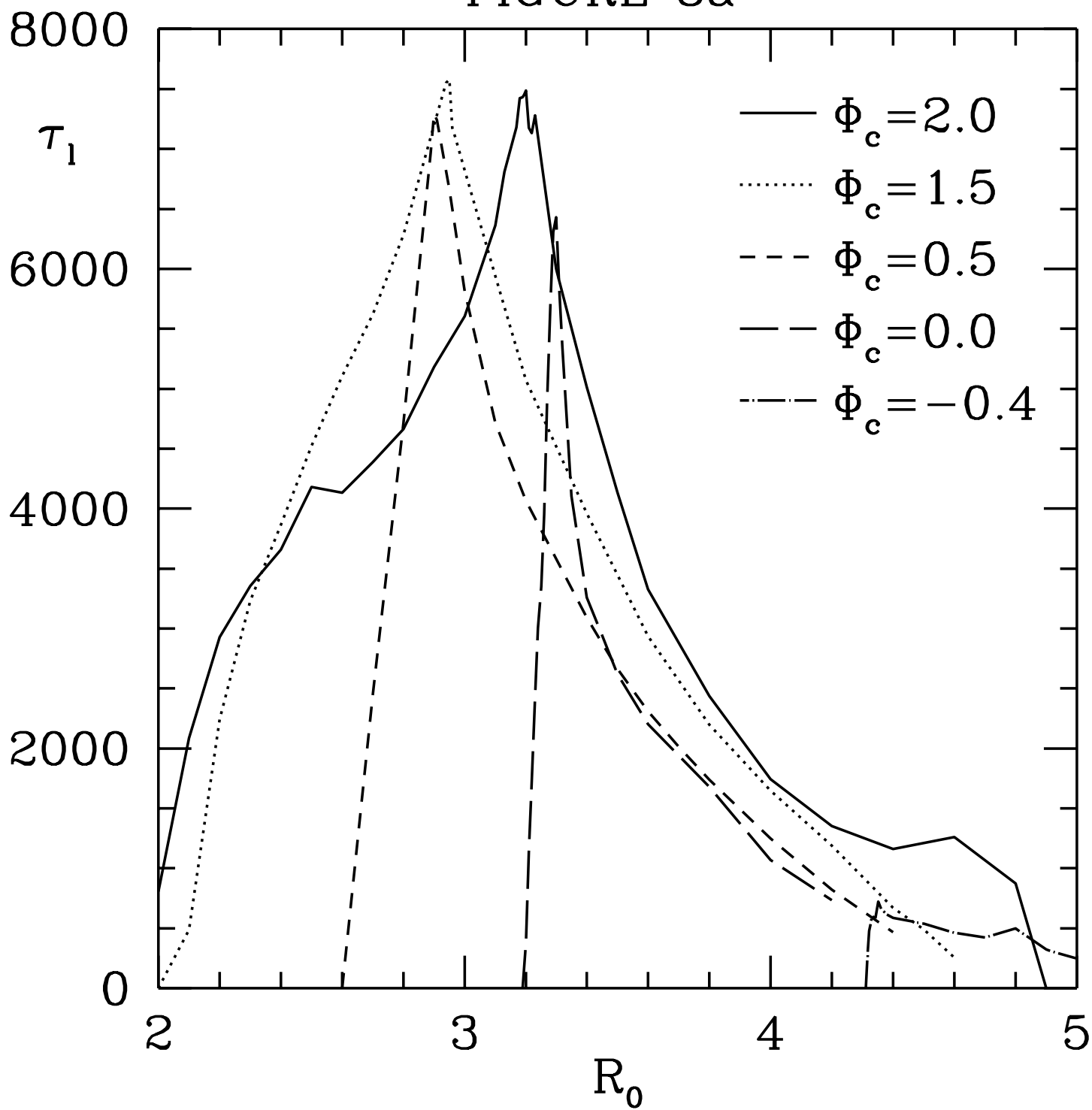
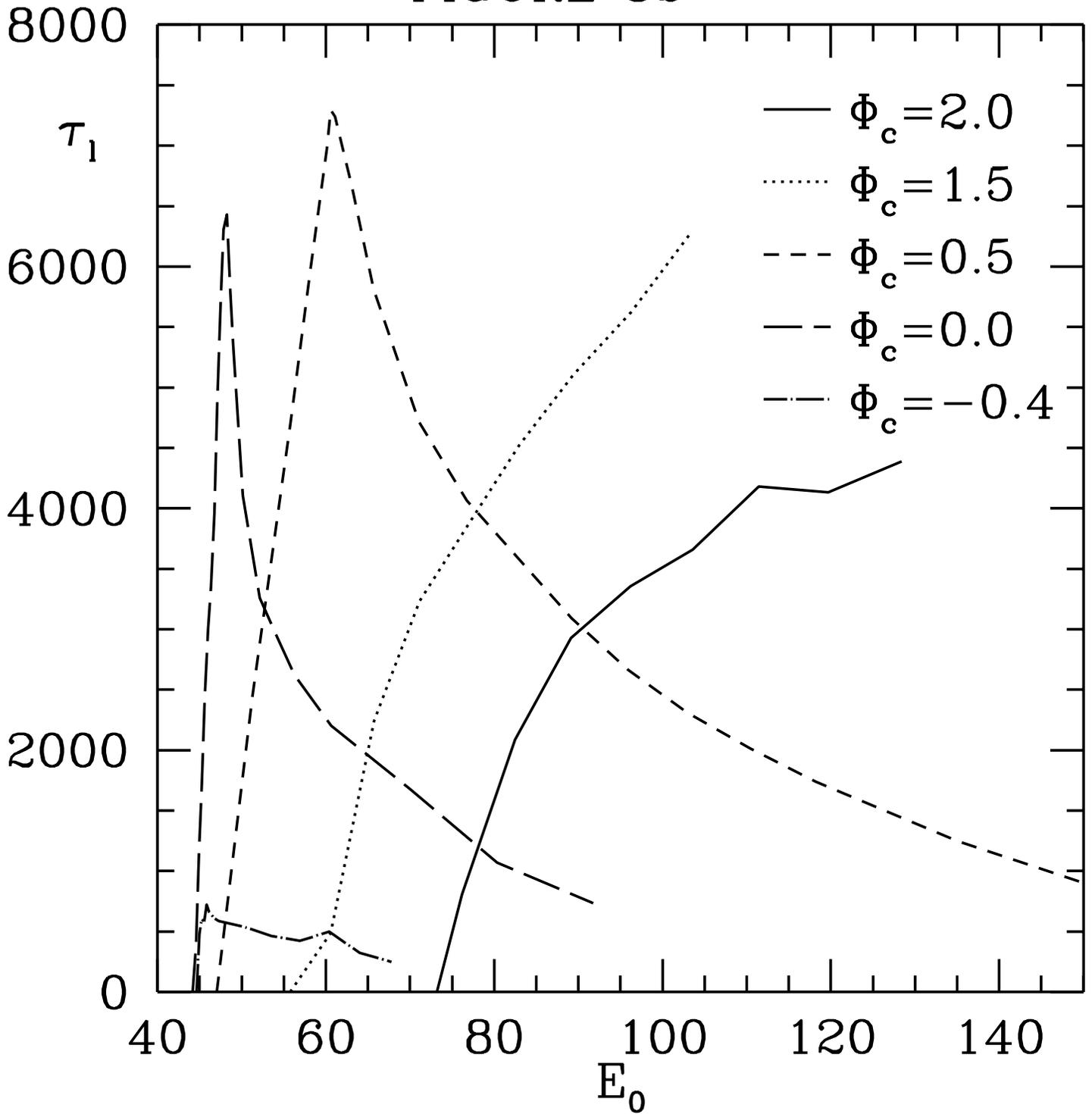
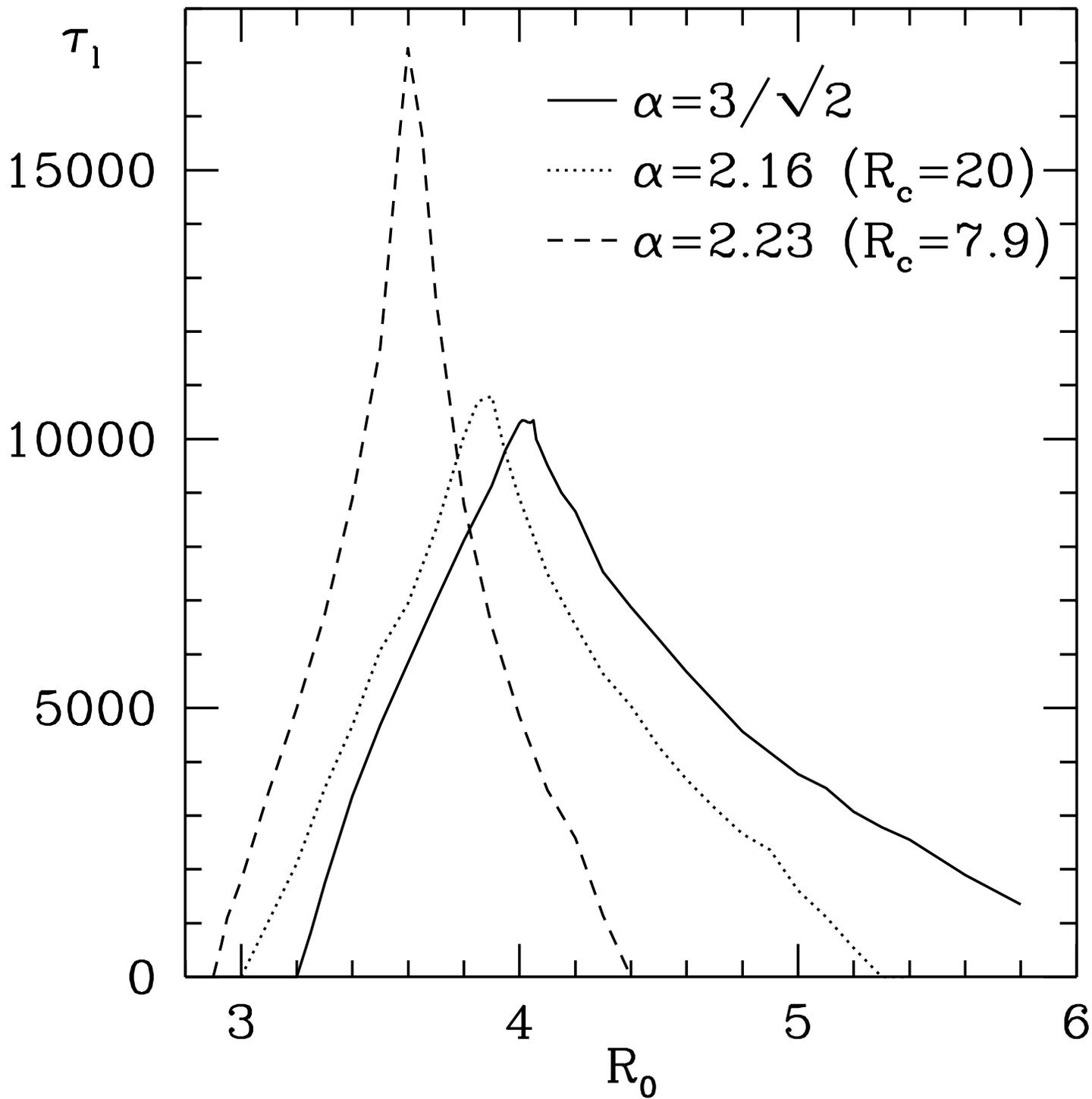
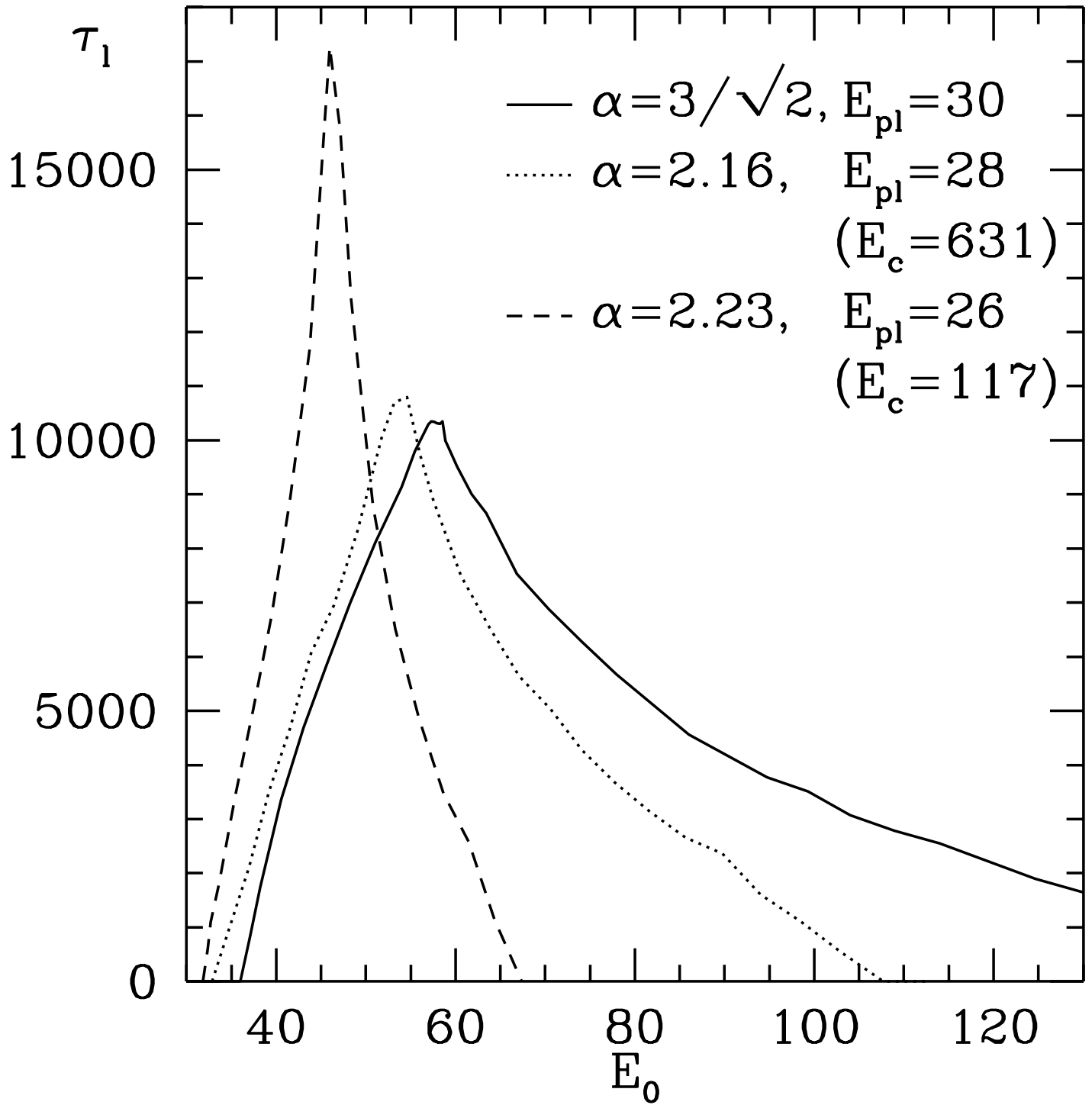
\end{document}